\documentclass[pdflatex,sn-mathphys-num]{sn-jnl}

\usepackage{graphicx}%
\usepackage{multirow}%
\usepackage{amsmath,amssymb,amsfonts}%
\usepackage{amsthm}%
\usepackage{mathrsfs}%
\usepackage[title]{appendix}%
\usepackage{xcolor}%
\usepackage{textcomp}%
\usepackage{manyfoot}%
\usepackage{booktabs}%
\usepackage{algorithm}%
\usepackage{algorithmicx}%
\usepackage{algpseudocode}%
\usepackage{listings}%
\usepackage{graphicx}
\usepackage[capitalise]{cleveref}
\usepackage{subcaption}
\usepackage{siunitx}

\title{Bridging the Gap on AI-Assisted Scientific Software Development Through Transparency and Traceability}
\author*{Chaitanya Bhave \thanks{Corresponding author}}
\email{chaitanya.bhave@inl.gov}
\author{Pierre-Cl\'ement A. Simon}
\author{Casey Icenhour}
\author{Lin Yang}
\author{Cody J. Permann}
\author{Daniel Schwen}
\author{Christopher S. Ritter}

\affil{\orgname{Idaho National Laboratory}, \city{Idaho Falls}, \state{Idaho}, \country{USA}}

\date{March 2026}

\begin{document}

\abstract{
The widespread adoption of AI-assisted development in scientific software is not a future concern---it is a present reality. Researchers are already using large language models to write code, generate test cases, and draft documentation, yet this practice remains largely unacknowledged and unguided in formal workflows and published work. This ad hoc, ungoverned use of AI represents a systemic risk to scientific software quality, particularly in safety-relevant modeling and simulation tools subject to strict Software Quality Assurance (SQA), or even Nuclear Quality Assurance Level 1 (NQA-1) standards, for which traceability, independent verification, and documented procedures are paramount. The question facing the scientific software community is, therefore, not whether to permit AI-assisted development, but how to govern it responsibly. This paper proposes guidance for AI-assisted code development in the context of strict software quality assurance. Using TMAP8---an open-source tritium migration code for fusion energy---as a demonstration platform, we propose a structured framework for AI-assisted verification and validation (V\&V) case development. V\&V case development represents the ideal proving ground for establishing that governance: because validation cases have known solutions, correctness is objectively measurable, errors are identifiable by design, and the artifacts are fully auditable. The proposed guidance, developed based on practical experience described herein, operates within NQA-1 requirements, preserves human accountability, and establishes the disclosure and review standards that responsible AI-assisted scientific software development demands.}

\maketitle

\section{Main}\label{sec:1_main}

The accelerating pace of scientific discovery offered by artificial intelligence (AI) represents an immense opportunity. 
Many of today’s most pressing technological challenges outstrip the capacity of traditional, sequential research models. 
AI models -- and especially, emerging agentic AI systems -- offer a powerful means to accelerate scientific discovery and technological breakthroughs through increased productivity, unparalleled large data analysis, and creative problem-solving. 
Yet such acceleration must be matched by uncompromising standards of quality, validation, and governance; without them, increased speed risks undermining reliability, reproducibility, and safety. 
Ensuring that AI-enabled scientific and technological advances maintain the highest levels of rigor is, therefore, essential to realizing scientific and technology breakthroughs at the pace and scale envisioned. 

This unprecedented context, and the rise of large language models (LLMs) in particular, has enabled a new paradigm in software development, where code is written by LLM ``agents'' instead of human developers, and the human role has shifted from code development to interacting with the agents for software engineering, planning, and code review \cite{sapkota2025vibecodingvsagentic,robbes2026agenticmuchadoptioncoding}. Modern AI systems can do far more than convert plain-language instructions into lines of code. They can interpret what a task actually demands, work with development tools directly, and continuously improve their output until the result is genuinely complex, functional, tested, and documented software. This new paradigm has been enabled by giving LLMs access to a set of tools — functions they can invoke to read, write, build, and run code directly in the development environment \cite{masterman2024landscape}.

As a result, software engineering now accounts for nearly half of all observed agentic AI activity \cite{anthropic2026agents}, and the duration of autonomous AI operation before human intervention has nearly doubled over a span of months \cite{anthropic2026agents}. The scientific community has not been insulated from this transition. Researchers are increasingly using LLMs to both write scientific code \cite{Ciriello2026, Kousha_AI_summary_2024} and to support or even carry out the entire scientific process \cite{zhang2025exploring, wang2024scipip, xie2025far, yamada2025ai, weng2025deepscientist, lyu2026evoscientist}. Examples include the agentic system developed by Yamada et al.~\cite{yamada2025ai} and the multi-agent system from Lyu et al.~\cite{lyu2026evoscientist}, which produced entirely AI-generated peer-reviewed papers. Despite the increasing use of agentic tools to write code, write documentation, and generate tests, these practices have remained largely unacknowledged, and the scientific community lacks a clear understanding of AI's strengths and limitations when it comes to formal workflows and published works \cite{Ciriello2026}. 

The consequences of this silence and lack of quality control are already visible. An analysis of papers accepted to the Thirty-Ninth Annual Conference on Neural Information Processing Systems (NeurIPS 2025) found more than 100 hallucinated citations across 53 papers, each missed by authors and multiple peer reviewers \cite{nazar_shmatko_2026}. In scientific software development, the analogous failure is not a fabricated reference but a subtly incorrect implementation: incorrect unit conversions, flawed formulae, or logically inconsistent code that appears plausible. Recent work has shown that LLMs frequently generate code containing outputs that appear correct but contain semantic or logical errors, which are  referred to as  ``hallucinations''\cite{lee2025hallucinationcodegenerationllms,zhang2025llmhallucinationspracticalcode}. Empirical studies further indicate that AI-generated code introduces significantly more defects, vulnerabilities, and maintenance issues than human-written code \cite{cotroneo2025humanwrittenvsaigeneratedcode}. 

These new AI-assisted development workflows challenge core assumptions underlying traditional software quality assurance (SQA) principles. In many cases, both implementation and testing may be produced by similar agentic models, raising concerns about correlated failure modes rather than independent verification. At the same time, AI-assisted development can increase code volume and complexity, increasing the burden of thorough reviews \cite{huang2026hallucination}.

Despite these risks, there is currently no standardized framework for applying SQA principles to agentic AI-generated code. This gap is particularly significant in safety-critical scientific domains, where existing assurance practices were developed under the assumption of human authorship and independent verification.

The instinct to respond with prohibition is understandable but counterproductive. Restricting AI use in scientific workflows will not eliminate the practice; it will drive it underground, producing the ad hoc, ungoverned, undisclosed use that poses great risk to trust in the quality of code development and, accordingly, its potential output \cite{Ciriello2026, Tang2025}. The productive question is not whether to permit AI-assisted development but how to govern it responsibly: how to make AI involvement visible, traceable, and subject to appropriate quality controls without impeding the genuine productivity gains that motivate its adoption \cite{roychoudhury2025agenticaisoftwareengineers}.

Standards bodies in safety-critical domains have already developed guidance to ensure high software quality for human developers. In the nuclear fission energy space, the American Society for Mechanical Engineers (ASME) has developed the Nuclear Quality Assurance Level 1 (NQA-1) standard~\cite{ASMENQA12017}. This standard establishes requirements for quality assurance programs across the nuclear industry and defines frameworks for planning, controlling, and executing activities related to nuclear facilities in order to ensure safety and reliability. The NQA-1 standard contains SQA requirements for software used in the nuclear fission industry. In this set of requirements (around 150 in total to be ``compliant''), a graded approach is followed (i.e., more rigorous controls for safety-critical software and lower, or no, requirements for non-safety software). NQA-1-derived SQA focuses on the development process itself and ensures that the process is documented; that the objectives of the code are tracked with requirements, documentation, and testing; that the people implementing and maintaining the code are qualified; and that all components of this process are traceable, among many other things. Note that while not every scientific field relies on the NQA standard, it is used here as a reference for the AI-governance consideration. Within this frame of reference, we can explore AI-in-the-loop SQA that is also standards compliant in a field that, by necessity, must embrace high levels of quality.

Verification and validation (V\&V) case development offers a natural proving ground for experimenting with AI-appropriate software quality control. V\&V cases exercise a simulation code against problems with known analytical solutions (verification) or experimental measurements (validation) and are the established mechanism by which computational science standards demonstrate software correctness \cite{Oberkampf_Roy_2010, IEEE1012,Jakeman2025_VV}. Unlike general production code, V\&V cases possess objective, measurable correctness criteria: a V\&V case either reproduces its reference solution within specified tolerances or it does not. This makes errors detectable by construction, regardless of whether they were introduced by a human or an AI agent. In contrast, ``correctness'' within NQA-1 is defined a little differently. Independent evidence-based review is used to establish acceptance against the defined criteria, where ``correctness'' is judged through the lens of the end use. Common to both systems, however, is a structured approach to the end goal. Well-defined and well-executed V\&V cases provide high levels of documentation and implementation quality to allow for good reproducibility. Thus, the same properties that make V\&V cases effective for catching human coding errors (e.g., requirements, documentation, testing) make them well suited for evaluating and governing AI-generated code, a connection that has not yet been systematically explored. The Tritium Analysis Migration Program, version 8 (TMAP8) \cite{Simon2025}, a leading open-source NQA-1-compliant fusion energy simulation tool (even if fusion industry tools are not required to be NQA-1 compliant) based on the Multiphysics Object Oriented Simulation Environment (MOOSE) \cite{harbour2025moose}, provides the concrete domain in which this connection is developed.

Most fusion power plant conceptual designs rely on the fusion of deuterium (D) and tritium (T) to generate energy \cite{Freidberg2007,Ongena2016}.
While deuterium is naturally abundant in seawater, tritium is scarce and radioactive, with a half-life of approximately 12.3~years \cite{Lucas2000}.
The current global, available tritium inventory, which is primarily produced as a byproduct of heavy-water Canada deuterium uranium (CANDU) fission reactors, is estimated to be only approximately 25--30~kg \cite{PEARSON20181140,NASEM2021}.
However, a single 1~GW fusion power plant is expected to consume on the order of 55~kg of tritium per year \cite{Abdou2021,NASEM2021}, a demand that far exceeds available supplies.
This stark imbalance will force fusion energy systems to breed and process tritium on site primarily through the interaction of fusion-produced neutrons with lithium in the tritium breeding blanket~\cite{Abdou2021,Kovari2018}.
Unfortunately, tritium permeates through surrounding structural materials, leading to tritium retention and losses, as well as undesirable local radioactivity, leading to safety concerns.
Managing tritium inventory and predicting transport through complex, often irradiated material systems is, therefore, one of the central challenges to the deployment of safe and economically viable fusion energy \cite{Causey2002,SHIMADA2020251,FESAC2018TEC,FESAC2020,DOESAC2015,NASEM2021,DOEFusionRoadmap2025}.

TMAP8 addresses this challenge by enabling multiscale, multiphysics simulations of tritium transport \cite{Simon2025,Simon2026,Franklin2025,Shimada2024,yang2026elucidating} in fusion energy, where the opportunities and risks described above are particularly pronounced. 
It has been developed to be NQA-1 compliant and proposes a growing open-source list of more than 38 V\&V cases to demonstrate its accuracy \cite{Simon2025,Simon2026}. This library of V\&V cases makes TMAP8 the ideal proving ground for AI governance frameworks.

The United States and other nations have set ambitious goals for producing abundant, reliable energy, the timelines of which demand concurrent, unprecedentedly rapid scientific breakthroughs and technological advancements, in part through scientific computing. Meeting these ambitions while ensuring software quality will require a clear, deployable, and adaptable governance framework for AI-assisted code development. This paper presents a governance framework for AI-assisted code development in NQA-1-governed open-source scientific software, demonstrated through two new validation cases for TMAP8.

The framework specifies how AI involvement should be disclosed and recorded in commit metadata, how review requirements should scale with the degree of AI involvement, how automated quality gates can be designed to detect silent AI errors in V\&V contexts, and how human accountability is preserved throughout. 
These requirements are encoded in a lightweight, repository-level AGENTS.md specification \cite{gloaguen2026evaluating}, enabling practical adoption within NQA-1-governed (or any high SQA-conscious) open-source projects. In addition to providing a governance framework, in this paper we detail the thought process that went into designing the framework and how it was tested through the development of two validation cases, and we share the improvements to the governance process that we believe should be addressed in future work. Thus, we attempt to provide an iterative approach to developing governance frameworks for AI-assisted code development systems that are constantly evolving. 

The framework rests on a straightforward premise: the transition to agentic development is already here \cite{anthropic2026agents}. The relevant choice is therefore not between AI-assisted and human-only development, but between governed and ungoverned AI assistance. V\&V cases, with their objective correctness criteria and bounded scope, provide a low-risk, high-signal environment in which to establish governance practices that can later extend to broader scientific software development.

\section{Identified best practices}
\label{sec:2_methodology}

The use of agentic AI systems in scientific software development introduces failure modes that are not explicitly addressed by existing SQA practices. In particular, LLMs can produce outputs that are syntactically correct and contextually plausible but semantically incorrect. Additionally, LLMs can hallucinate interfaces between codes. These errors are often difficult to detect through informal review alone and may propagate if not systematically constrained. To address this, we adopt a set of design principles that preserve the core objectives of SQA. These objectives include verifiability, traceability, independence of review, and accountability. At the same time, the principles accommodate the capabilities and limitations of AI-assisted workflows. In this work, AI assistance refers specifically to agentic code generation systems operating within development workflows rather than passive or advisory use.

Verifiability is anchored in V\&V cases, which provide known analytical or experimental reference solutions against which all intermediate and final outputs are evaluated. We distinguish two uses of the term throughout: in the V\&V sense, verification confirms that a computational model accurately represents its underlying mathematical formulation; in the SQA sense, it refers to confirming that a software artifact meets its specified requirements. Both apply here. All contributions, regardless of origin, are exercised against the full V\&V suite through automated continuous integration (CI) pipelines.

Because LLMs exhibit non-deterministic behavior, they require bounded operating conditions. AI systems are constrained to well-defined task scopes with explicit expectations for acceptable outputs, and human intervention is required when outputs fall outside predefined criteria or when verification results are inconclusive. The context provided to the model is deliberately controlled: too little risk degenerates outputs; too much increases hallucinations.

As we work through the remainder of this section, we will highlight specific requirements from the ASME NQA-1 specification~\cite{ASMENQA12017}, where those requirements are addressed in the Software Quality Assurance Plan that governs software development practices for MOOSE and MOOSE-based applications (PLN-4005~\cite{PLN-4005}), and how we are approaching those requirements and governance practices within our agent-focused development activities. \cref{tab:nqa1_mapping} summarizes these governance practices alongside the specific NQA-1 sub-clauses and PLN-4005 sections each practice satisfies, and makes the compliance argument explicit and auditable.

PLN-4005 (Sections~6.2.1 and~9.4.1), the Software Quality Assurance Plan that governs software development practices for MOOSE and MOOSE-based applications, states that software verification must be performed by an independent human reviewer. This practice is consistent with NQA-1 SP~2.7-402.1. No agentic system, including adversarial review agents, is considered sufficient to satisfy this requirement. The role of AI-assisted checks is therefore limited to improving the quality and consistency of submitted artifacts prior to human review, rather than replacing it. The work presented here reinforces that principle in the context of AI-assisted workflows.

All contributions must carry metadata documenting the degree and nature of AI involvement, and they must be linked to a corresponding issue describing the intent, design rationale, and expected impact of the change. This practice satisfies the formal documentation requirements of NQA-1 SP~2.7-203.2(c) and SP~2.7-401, which are reflected in PLN-4005 Section~9.8.2.1. These attributes must be recorded at the commit level, and this effort must include an associated human-readable log of AI interactions. This practice ensures traceability across requirements, implementation, and validation within the version control system. These records are maintained within the repository, enabling developers to stochastically reconstruct the development process from repository artifacts using current versions of AI tools. Pre-commit checks enforce formatting, metadata completeness, and adherence to contribution guidelines at the point of submission, consistent with the configuration management requirements of NQA-1 SP~2.7-203 and the standards documentation requirements of SP~2.7-500 and PLN-4005 Section~9.8.2.3, reducing the likelihood that invalid or incomplete contributions will enter the review process.

Contributions that fail V\&V or do not satisfy established provenance requirements are rejected and must be revised prior to review or integration, consistent with NQA-1 SP~2.7-202 and SP~2.7-404.4 and PLN-4005 Sections~9.4.1 and~9.8.2.4. While adversarial review agents may be employed to critically assess generated outputs and may provide a supplementary quality layer, they were not required in this work and do not serve as substitutes for required human oversight. Contributions are evaluated not just for correctness but for genuine utility. Code that increases volume without improving functionality, clarity, or testability is treated as a defect and subject to revision or rejection. These constraints are particularly important in AI-assisted workflows, where code can be generated at a rate that exceeds human review capacity.

\begin{table}[htbp]
\centering
\caption{Mapping of AI-assisted development governance practices to NQA-1 sub-clauses and PLN-4005~Rev.~10 sections.}
\label{tab:nqa1_mapping}
\small
\begin{tabular}{p{0.47\textwidth}p{0.24\textwidth}p{0.19\textwidth}}
\hline
\textbf{Governance Practice} & \textbf{NQA-1 Sub-clause} & \textbf{PLN-4005 Section} \\
\hline
Commit-level AI disclosure metadata (degree and nature of involvement) & SP~2.7-203.2(c) & \S9.8.2.1 \\
\hline
Issue-linked commits (intent, design rationale, CI impact) & SP~2.7-401 & \S9.8.2.1 \\
\hline
Session logs in version control & SP~2.7-402.1; SP~2.7-404.3 & \S5.3 \\
\hline
Human independent reviewer required; no agentic system is sufficient& SP~2.7-402.1 & \S6.2.1; \S9.4.1 \\
\hline
\texttt{AGENTS.md} specification version-controlled with codebase & SP~2.7-500 & \S10.1 \\
\hline
Pre-commit hooks enforcing non-negotiable provenance requirements & SP~2.7-203; SP~2.7-500 & \S9.8.2.3 \\
\hline
Full V\&V test suite executed for all contributions via CI pipeline & SP~2.7-404; SP~2.7-404.4 & \S9.5.2; \S9.4.1 \\
\hline
Rejection of contributions failing V\&V or provenance checks & SP~2.7-202 & \S9.4.1; \S9.8.2.4 \\
\hline
Human developer as author bearing full accountability & Part~I R3-401 & \S9.4.1 \\
\hline
\end{tabular}
\end{table}

These practices are encoded as enforceable repository-level policy through a governance specification file (\texttt{AGENTS.md}), which defines requirements for task execution, validation, documentation, and review and is version-controlled alongside the codebase. By embedding compliance requirements directly in the development infrastructure rather than relying on developer discretion, the framework ensures that AI-assisted contributions are held to the same standards as human-authored code. This approach extends, rather than replaces, existing SQA processes, ensuring alignment with established requirements for traceability, reproducibility, and independent verification. The following section describes how this framework was applied in practice.

\section{Demonstration of validation case development with agents}
\label{sec:validation}

In this section, we discuss using the guidance provided in \cref{sec:2_methodology} to implement two new validation cases in TMAP8 using coding agents.
The two cases are based on previously published experimental papers that were carefully selected using the following criteria:

\begin{enumerate}
    \item The experimental studies were selected from external institutions to reflect realistic development conditions, where both the developer and the agent rely solely on published information.
    \item The validation cases were required to be distinct from existing TMAP8 V\&V cases, ensuring that each case introduced at least one novel modeling component.
    \item One case was selected from a study that provides an established model that accurately captures the experimental data, representing an \textit{implementation} challenge focused on correct translation into TMAP8.
    \item The other case was selected from a study without a published model and thus required hypothesis-driven model construction. This represents a \textit{scientific reasoning} challenge involving the identification of relevant mechanisms, formulation of candidate models, and iterative refinement against experimental data.
\end{enumerate}

Based on these criteria, we selected a publication by Kobayashi et al.\ on tritium release from neutron-irradiated Li$_2$TiO$_3$ \cite{kobayashi2015developing} as the first case. TMAP8 previously offered no validation cases on irradiated lithium-based ceramics nor cases involving the annealing of irradiation-induced defects. Critically, Kobayashi et al.\ proposed an analytical model in the same paper to capture the experimental tritium release behavior, providing a clear modeling target that the agent could implement and compare against (criterion n$^{\circ}$3).

The second case was drawn from a paper by Kremer et al.\ \cite{Kremer2022oxide}, which reports experimental results of deuterium release from self-irradiated tungsten with thin native oxide films (5 to 100~nm). Similarly, no prior TMAP8 validation case had modeled the influence of surface oxide layers on hydrogen isotope retention and release. To the best of the authors' knowledge, no researchers have developed a model to reproduce the reported experimental data and published their results. This case constitutes a \textit{scientific reasoning} challenge that requires an agent to construct a hypothesis-driven model rather than directly implementing an existing one (criterion n$^{\circ}$4). The agent must (i) identify relevant physical mechanisms and hypothesize how they interact, (ii) draw on analogous TMAP8 cases to inform the model structure while recognizing that the oxide-layer physics fall outside TMAP8's established V\&V envelope, and (iii) systematically compare competing model formulations against the experimental data to identify the most physically consistent and predictive description. This iterative, hypothesis-driven development process represents a qualitatively different and higher challenge than faithful re-implementation of a known model.

Together, these two cases provide complementary testbeds for evaluating AI-assisted development across distinct cognitive regimes.

\subsection{Validation case 1}\label{sec:validation:case_1}

The first validation case models tritium thermal desorption spectroscopy (TDS) from neutron-irradiated Li$_2$TiO$_3$ \cite{kobayashi2015developing}. The experimental data exhibit temperature-dependent release behavior influenced by trapping at irradiation-induced defects and defect annihilation at higher temperatures. A published mechanistic model is available for this system, providing a well-defined target for implementation within TMAP8. This case therefore represents an \textit{implementation} challenge, where the agent is tasked with translating and verifying an existing model. The details of the model formulation and parameters are provided in the Supplementary Materials.

This validation case was developed using a Claude-code agent-assisted workflow with an Opus 4.6 LLM backend and was focused on reproducing an existing published model. At session initialization, the agent was provided with the repository-level \texttt{AGENTS.md} specification, which defines requirements for task execution, testing, documentation, and provenance \cite{tmap8-pr-397}. The developer prompted the agent to read the experimental paper and respond to a structured set of questions that guided the implementation process:

\begin{itemize}
    \item What experimental observations and data are reported?
    \item What physical mechanisms are likely to govern the observed behavior?
    \item How could these mechanisms be represented within the TMAP8 modeling framework?
    \item Propose a concrete plan (\texttt{/plan}) to implement and evaluate candidate models.
\end{itemize}

These prompts were intended to ensure that the agent correctly interpreted the published model before generating implementation artifacts. As required by the \texttt{AGENTS.md} workflow, the agent requested an issue number before proceeding, ensuring that the resulting changes were linked to the corresponding development record.

The agent then supported two stages of development. In the first stage, it extracted the governing equations and parameters from the source paper, mapped the model formulation onto existing TMAP8 constructs, generated input files and test artifacts, and drafted documentation consistent with project conventions. In the second stage, after preliminary comparison with the experimental TDS spectrum, the agent assisted in configuring and executing a Bayesian calibration workflow using the MOOSE stochastic tools module, which was not part of the original paper by Kobayashi et al. \cite{kobayashi2015developing}. This included preparing parameter ranges, formatting optimization input files, generating scripts for post-processing and plotting, and updating comparison figures after optimization. The generated input files and session file logs (\texttt{.jsonl}) were committed and merged into the TMAP8 codebase~\cite{tmap8-pr-402}.

\begin{figure}[htbp]
    \centering\includegraphics[width=0.7\linewidth]{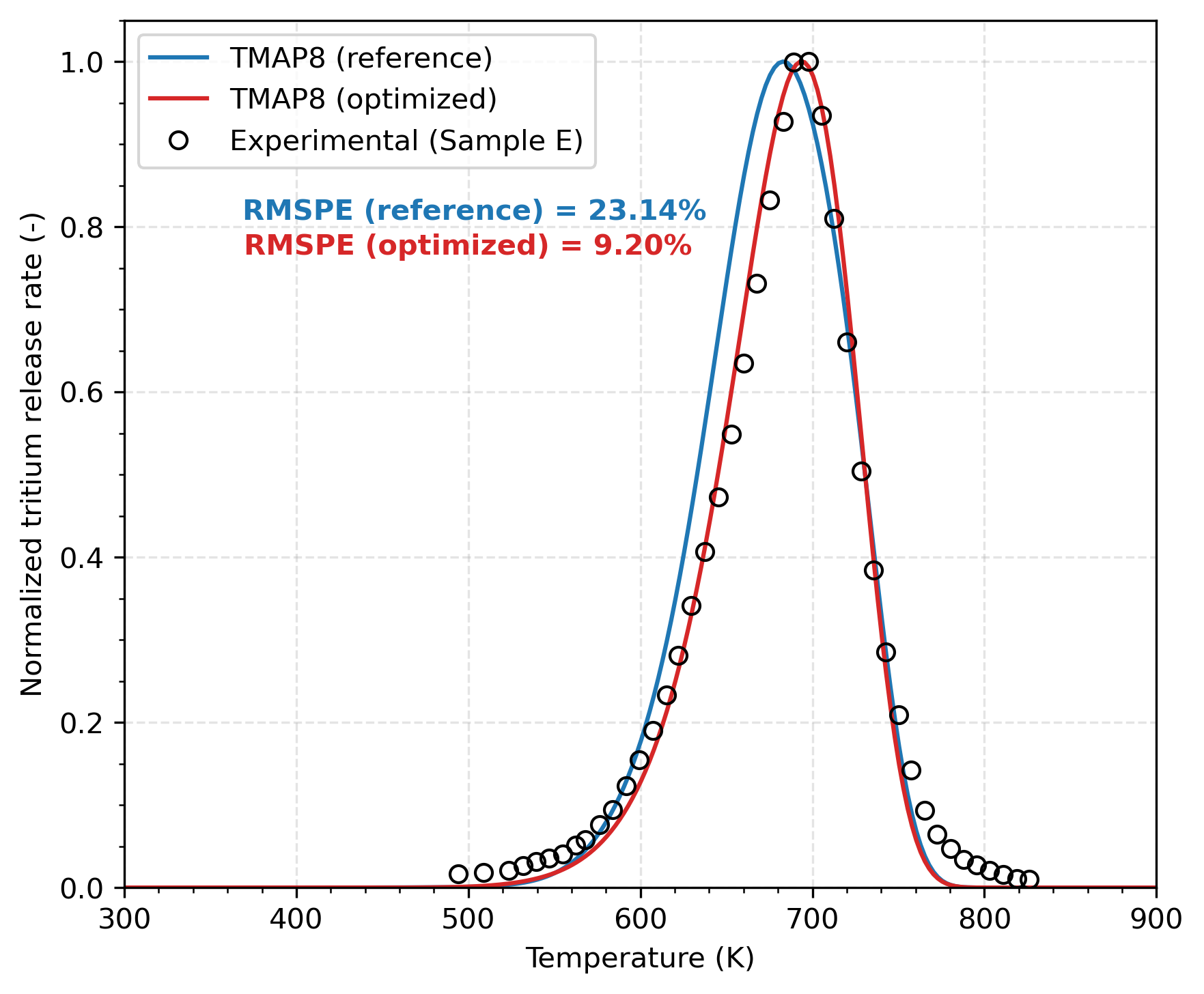}
    \caption{Comparison of TMAP8 calculation with the experimental TDS data for Sample~E (high defect density). Experimental data from Kobayashi et al.~\cite{kobayashi2015developing}. The preliminary model has a root mean square percent error (RMSPE) of 23.1\%. Bayesian-optimization of parameters against the experimental TDS data reduces this RMSPE to 9.2\%.}
    \label{fig:val1:comparison}
\end{figure}

As shown in \cref{fig:val1:comparison}, the resulting implementation reproduces the experimentally measured normalized tritium release rate, demonstrating that the agent-assisted workflow can correctly translate established physical models into executable code. Further details of the model formulation and parameterization are provided in the Supplementary Materials.

The workflow accelerated several routine but time-consuming tasks, including input-file formatting, test generation, gold-file updates, plotting, and \texttt{MooseDocs} documentation \cite{Slaughter03072021}. However, it also exposed characteristic failure modes. For example, the agent initially omitted the defect-annihilation term in the trapping equation, mirroring an ambiguity in the original publication. This omission led to an overestimation of trapped tritium and an incorrect sensitivity to defect annealing and was identified only through human review and testing. The agent also occasionally proposed unnecessarily complex implementation paths, such as a custom C++ post-processor output class (i.e., a MOOSE object that computes and outputs scalar quantities from a simulation) for the Bayesian objective function, when a simpler input-file-level solution was sufficient.

Overall, this case demonstrates that agent-assisted workflows can accelerate both the implementation of established models and subsequent calibration workflows, but only when bounded by explicit repository-level guidance and coupled to a human review capable of identifying physically plausible but incorrect implementations.

\subsection{Validation case 2}\label{sec:validation:case_2}

The second validation case is based on thermal desorption spectroscopy measurements of deuterium release from self-irradiated tungsten with thin oxide layers \cite{Kremer2022oxide}. The corresponding experiments probed the impact of oxide layers on hydrogen isotope retention and release, a key factor in interpreting laboratory measurements and relating them to fusion-relevant conditions. The details of the validation case are provided in the Supplementary Materials.

This validation case was developed using the Codex tool with a ChatGPT 5.4  LLM backend. In contrast to the first case, no published model was available to reproduce the experimental data, so the agent had to engage in hypothesis-driven model development rather than direct implementation.

At session initialization, the agent was provided with the repository-level \texttt{AGENTS.md} specification~\cite{tmap8-pr-397}, which defines constraints on task execution, documentation, testing, and provenance requirements.

The developer then prompted the agent to read the experimental paper and answer a structured set of questions designed to guide its reasoning process:

\begin{itemize}
    \item What experimental observations and data are reported?
    \item What physical mechanisms are likely to govern the observed behavior?
    \item How could these mechanisms be represented within the TMAP8 modeling framework?
    \item Propose a concrete plan (\texttt{/plan}) to implement and evaluate candidate models.
\end{itemize}

As in the first validation case, these prompts were intended to verify comprehension of the experimental system, but in addition they were meant to elicit hypotheses regarding the underlying physics and possible modeling strategies. The agent was encouraged to identify multiple plausible mechanisms (e.g., trapping, diffusion barriers, surface reactions) and to propose corresponding model formulations.

As required by the \texttt{AGENTS.md} workflow, the agent requested an issue number before proceeding, ensuring traceability of all subsequent work. Once the agent had been provided an issue identifier, it performed the following steps:

\begin{itemize}
    \item Extract key experimental features and constraints from the source paper;
    \item Identify relevant existing TMAP8 validation cases to inform model structure;
    \item Propose two candidate model formulations that incorporate the hypothesized physical mechanisms for the more novel part of the model;
    \item Generate input files, scripts, and tests to implement these formulations;
    \item Iteratively refine the model by comparing simulation outputs against experimental data.
\end{itemize}

A key aspect of this workflow was that it gave the developer the ability to rapidly prototype and compare alternative modeling approaches. For example, multiple representations of the oxide layer (the main novelty of this model) were explored and evaluated within a single development cycle, allowing the developer to assess their relative performance and physical consistency. This iterative, hypothesis-driven loop—comprising model proposal, implementation, evaluation, and refinement—was central to the development of this validation case. 
Being able to quickly implement new model formulations is key to scientific discoveries. Hypotheses can be quickly formulated, models can be rapidly deployed to test these hypotheses, and conclusions can be drawn on an accelerated timescale. In this case, for example, the oxide layer was first represented as a separate domain from the tungsten, with an interface between the two regions. This was the approach recommended by the model, but it complicated the model formulation more than necessary, which was quickly identified after the agent's initial development. A continuous approach, described in more detail in the Supplementary Materials, reduces complexity while capturing the main phenomena required for this case. The shift from the first approach to the second one was very rapid thanks to the agent and quickly delivered encouraging results.

Throughout this process, all agent interactions were logged in machine-readable \texttt{.jsonl} session files and linked to version-controlled commits, ensuring full traceability of the development history. These session files are linked in the pull request \cite{tmap8-pr-409}. This iterative process resulted in approximately 70 commits (some generated by humans only and others the result of collaboration between developer and agent), during which multiple candidate formulations were explored and refined within short iteration cycles. 

\begin{figure}[h]
\centering
\begin{subfigure}{0.48\textwidth}
    \centering
    \includegraphics[width=\textwidth]{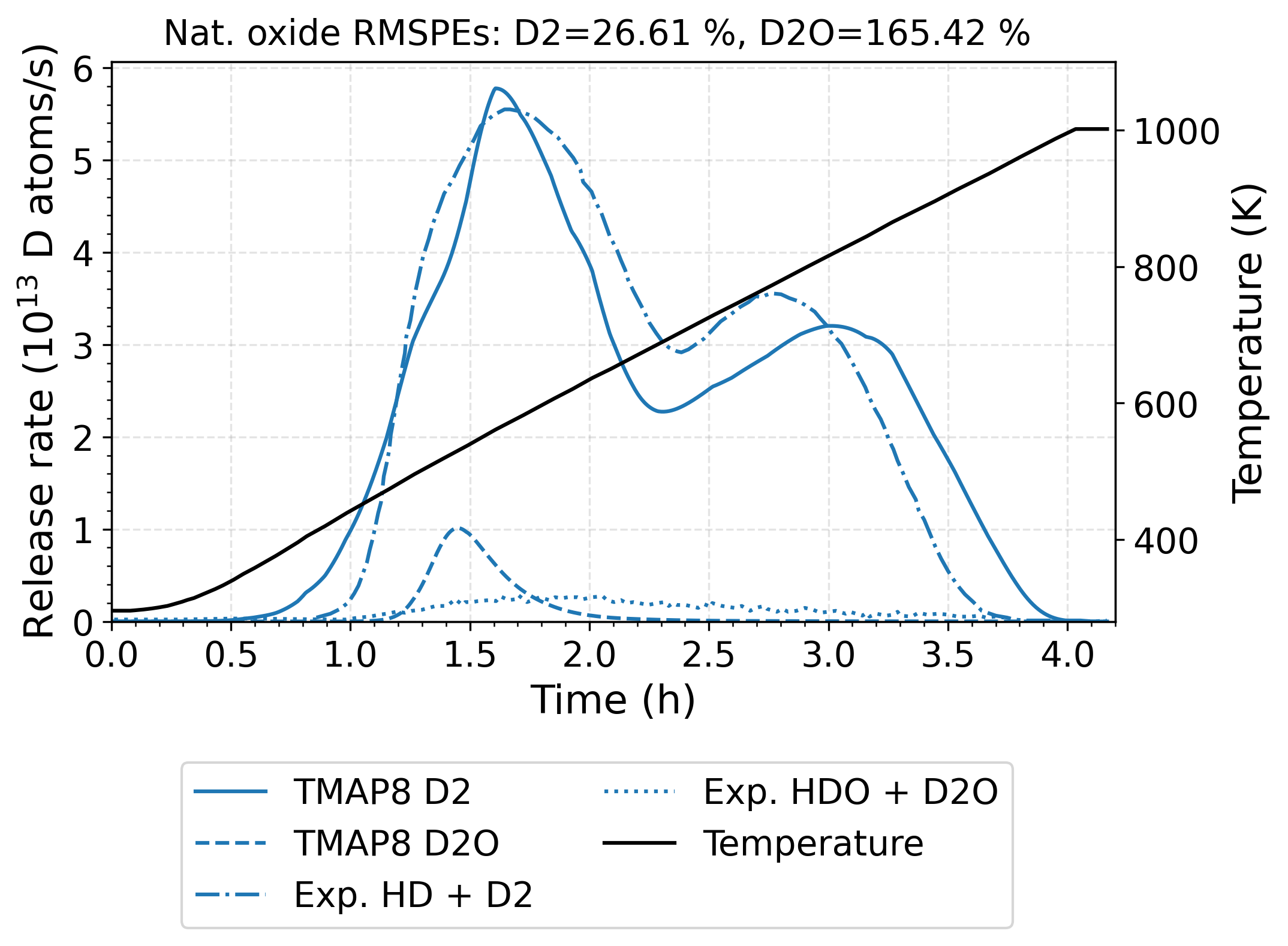}
    \caption{Natural oxide (1 nm)}
    \label{val-2k_natural_oxide_case_comparison}
\end{subfigure}
\hfill
\begin{subfigure}{0.48\textwidth}
    \centering
    \includegraphics[width=\textwidth]{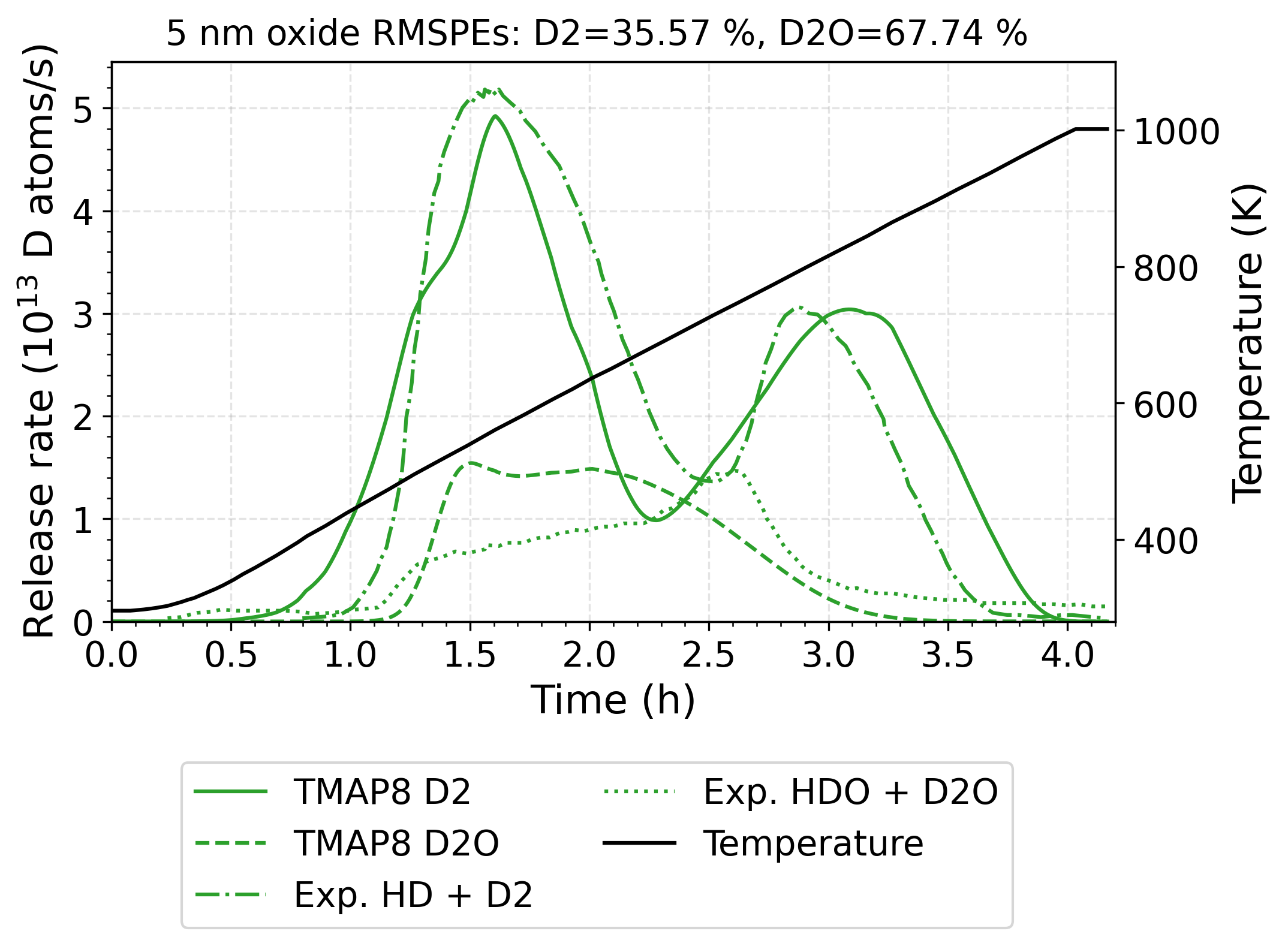}
    \caption{5 nm oxide}
    \label{val-2k_5nm_oxide_case_comparison}
\end{subfigure}

\vspace{0.5em}

\begin{subfigure}{0.48\textwidth}
    \centering
    \includegraphics[width=\textwidth]{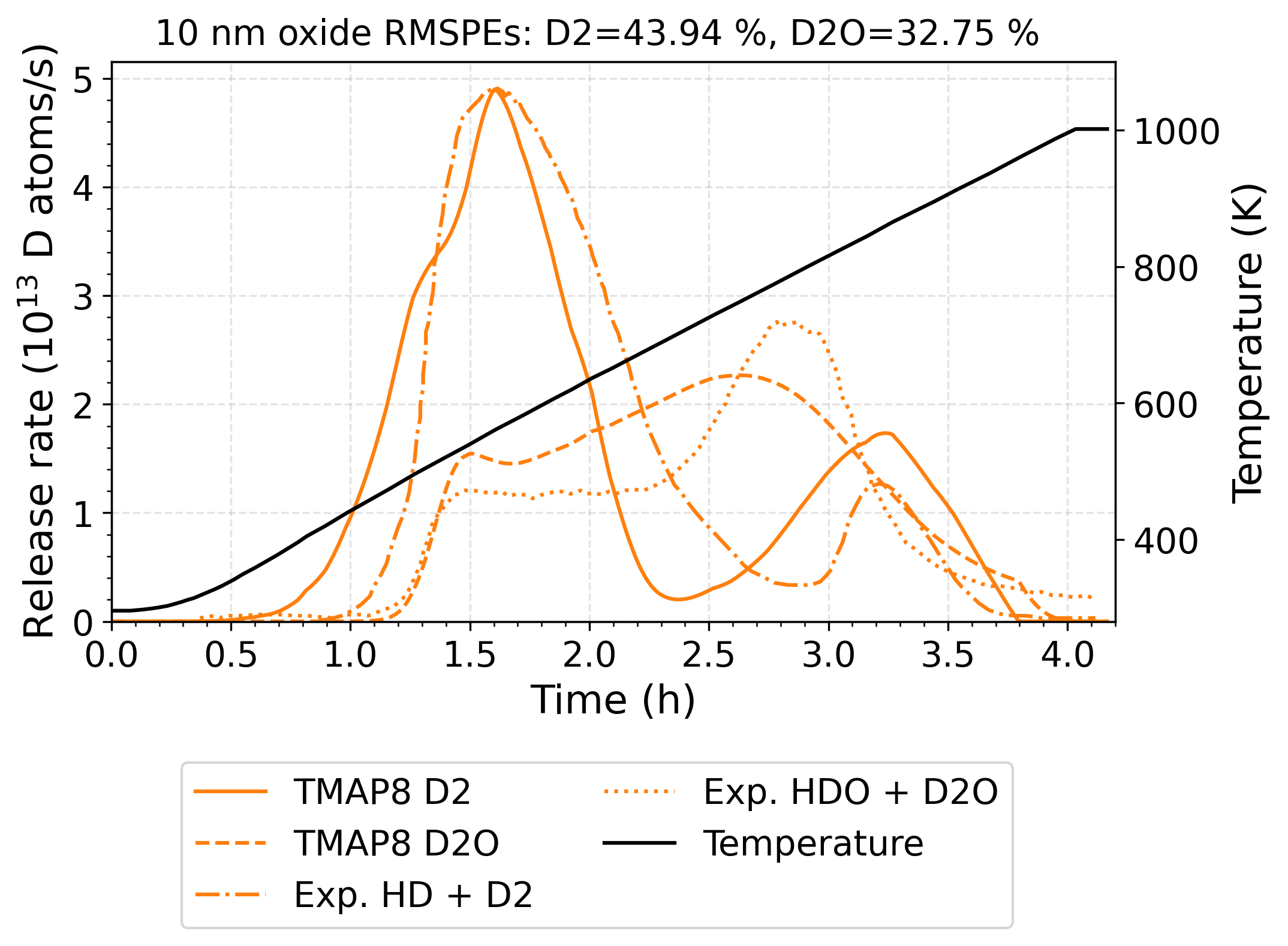}
    \caption{10 nm oxide}
    \label{fig:val-2k_10nm_oxide_case_comparison}
\end{subfigure}
\hfill
\begin{subfigure}{0.48\textwidth}
    \centering
    \includegraphics[width=\textwidth]{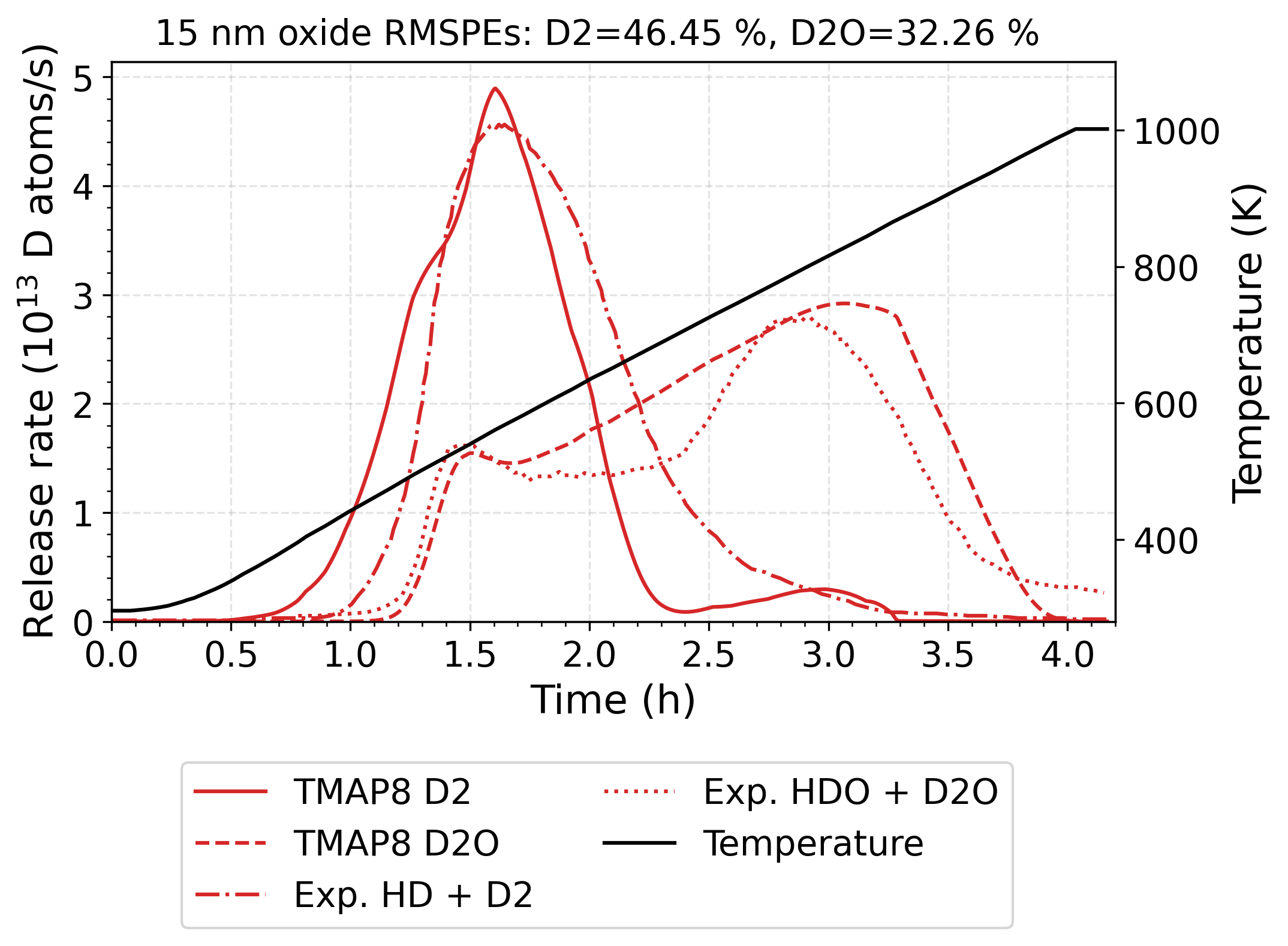}
    \caption{15 nm oxide}
    \label{fig:val-2k_15nm_oxide_case_comparison}
\end{subfigure}

\caption{Comparison of the D$_2$ and D$_2$O release simulation predictions against TDS experimental measurements from Ref.~\cite{Kremer2022oxide} for different oxide thicknesses (1 [natural oxide], 5, 10, and 15 nm).}
\label{fig:val-2k_comparison}
\end{figure}

The details of the final model formulation and model parameter values are provided in the Supplementary Materials. \cref{fig:val-2k_comparison} compares the final model's deuterium release predictions for all four oxide-layer configurations against the digitized HD + D$_2$ and HDO + D$_2$O desorption data from Fig. 6 of \cite{Kremer2022oxide}. The figure also overlays the digitized temperature history from the TDS experiment.
The experimental release curves all come from digitized grouped measurements in the paper, while the simulation curves come from the present calibrated model and are reported as deuterium-atom release rates to match the plotted grouped signals. 

The model reproduces the primary experimental trends across oxide thicknesses, demonstrating that the agent-assisted workflow can generate physically plausible and testable models in the absence of a predefined formulation. While discrepancies remain in specific peak shapes and magnitudes, the agreement is sufficient to validate the proposed development workflow. The goal is to derive physical insight from a phenomenological model, rather than to establish a definitive mechanistic model.

\section{Lessons learned, updated best practices, and discussion}
\label{sec:4_lessons_learned}

The two validation cases developed in this work offer a concrete basis for evaluating the governance framework proposed in \cref{sec:2_methodology}, and for identifying practical refinements that are needed as AI-assisted workflows mature in safety-critical scientific software development. Several cross-cutting lessons emerged from this experience that extend and update the best practices described above.

\subsection{AI agents as accelerators, not autonomous developers}

Across both validation cases, AI agents dramatically accelerated mechanical development tasks — formatting input files, generating plots, writing documentation, producing test scaffolding, and writing documentation in \texttt{MooseDocs} format — while freeing the human developer to focus on physical reasoning, model formulation, and quality judgment. This division of labor proved most productive when the developer maintained active ownership of the scientific direction.

As an illustrative measure of the impact of agent-assisted development, we note that a comparable validation case previously required approximately 4–5 days of development effort by an experienced contributor. Using the agent-assisted workflow described here, the same class of task was completed in approximately 6 hours by a developer with limited prior familiarity with the specific physical model. While this comparison is not controlled and should not be interpreted as a rigorous benchmark, it provides a practical indication of the potential reduction in development time for well-defined implementation tasks. There is a future opportunity for human factor research to study and fully understand agent-assisted development impacts.

In the first validation case, the agent accelerated the Bayesian optimization process by automating parameter formatting, generating gold files for tests, and producing comparison plots with consistent styling. In the second, it was particularly effective at leveraging existing TMAP8 validation cases — specifically val-2f~\cite{Simon2026,Kadz2026} — reusing established formulations where appropriate while spontaneously identifying and implementing features not previously developed in TMAP8, such as the introduction of an explicit tungsten oxygen layer to model D$_2$O surface release.

The second validation case also illustrates a qualitatively different and underappreciated benefit of AI assistance beyond the faster execution of known tasks: it provided the developer the practical ability to explore a broader hypothesis space within realistic development timelines. Where no published model existed, the agent enabled rapid prototyping of competing model formulations for the oxide layer, allowing a direct performance comparison that would have carried substantial cost in a purely human workflow. The ability to try a wider range of approaches is transformative to scientific software development where exploration is a necessity.

This productivity gain, however, comes with a corresponding responsibility. When agents generate large volumes of code, inputs, and documentation quickly, the burden of review grows proportionally — for both the developer and independent reviewers. The agent's tendency to drift toward overcomplicated solutions — introducing unnecessary duplication and structure — was observed in both cases and limits model interpretability. To be as valuable as possible, a validation model should be as simple as it can be while capturing the main experimental behavior. When prompted to review and simplify its outputs, the agent was able to identify opportunities to do so, but this required deliberate developer intervention and could not be assumed to occur naturally. The agent's susceptibility to reasoning drift and excessive exploratory branching, even when a plan was first defined, placed a greater burden on the developer to actively track the status of the case using their subject-matter expertise and to define next steps. Thorough sanity checks are therefore essential to ensure the relevance of the generated results.

More broadly, this shift in the developer's role — from direct code and script authorship toward prompt engineering and output review — demands that greater attention be paid to model complexity, performance, and accuracy. The cognitive cost of many individual tasks is reduced; the cognitive cost of maintaining scientific oversight of the overall development process is not.

\subsection{Hallucination as a persistent and structurally important risk}
\label{sec:4_hallucination}

LLM hallucinations in this context did not manifest as obviously implausible outputs, but as syntactically valid, contextually plausible, and semantically wrong implementations — the most dangerous category of failure in scientific software. In the first validation case, the agent reproduced an omission from the original reference paper~\cite{kobayashi2015developing}, failing to include the defect annihilation term $-k_{dp\text{-}da}\,C_T$ in the trapped concentration equation (Eq.~\ref{eq:val1:trapping}). This term describes tritium release from annealing defects; its absence led to an overestimation of trapped tritium and a lack of sensitivity to annealing effects. This error was only identified through adversarial human review and extensive testing, not through automated checks alone. In the second validation case, the agent was asked to digitize the figure containing the experimental measurements and hallucinated part of the data. While some complex trends were identified, other data points were either ignored or made up. The agent, however, did not express doubt regarding its accuracy in this task. The figure digitization eventually had to be made by hand. 

This finding underscores a structural limitation of AI-assisted workflows: when an agent is asked to implement an existing model from a published description, it tends to faithfully reproduce both the correct and incorrect features of that description. Independent human review is therefore not merely a procedural requirement but an epistemic necessity because humans can reason beyond the source document. 

In both cases, the structured constraints imposed by \texttt{AGENTS.md} reduced hallucination frequency by limiting the agent's operational degrees of freedom through explicit coding standards, testing requirements, and commit procedures. However, hallucination was not eliminated. When instructions were insufficiently specific, the agent occasionally produced plausible but incorrect outputs — for example, proposing a custom C\texttt{++} post-processor to simplify the Bayesian objective function and a Python script to calculate annihilation effects where simpler input-file-level solutions were sufficient.

Continued refinement of the \texttt{AGENTS.md} specification — particularly the addition of critical but previously undocumented requirements, such as TMAP8's distinction between \texttt{light} tests (those that run in ${\sim}2$ seconds or less) and \texttt{heavy} tests (those that take longer than ${\sim}2$ seconds) — is therefore an ongoing process. \texttt{AGENTS.md} must be treated as a living artifact of the development infrastructure that is updated iteratively as new failure modes are identified. In this work, areas for improvement identified during development were documented in a GitHub issue \cite{tmap8-issue-408}.

\subsection{Context limitations require structural countermeasures}

A recurring practical failure mode was the agent neglecting instructions that were explicitly documented in \texttt{AGENTS.md}, not because the instructions were absent, but because they fell outside the effective attention window during extended development sessions. This is a known limitation of current LLM architectures~\cite{huang2026hallucination}, and the framework must account for it structurally rather than relying on developer vigilance alone. For example, the agent encountered more difficulty running tests in validation case 2 as the discussion size increased. 

Several mitigations are worth pursuing. First, \texttt{AGENTS.md} should be structured to distinguish instructions by criticality, so that the most consequential requirements appear prominently and are less likely to be de-prioritized under context pressure. Second, complex tasks should be decomposed across agent teams to reduce the context load on any single agent. Third — and most reliably — developers should use automated pre-commit hooks to enforce non-negotiable requirements at the point of submission rather than relying on the agent to remember them.

The distinction between agent-level instructions and repository-level pre-commit hooks is important here. Git pre-commit hooks operate outside the agent's decision loop entirely; for requirements such as committing the machine-readable \texttt{.jsonl} session log in every pull request, a hook that auto-copies and \texttt{git add}s the file before commit is more reliable than an instruction the agent may overlook. Agent-level hooks and \texttt{AGENTS.md} instructions remain valuable for guiding agent behavior, but non-optional provenance requirements are better enforced at the infrastructure level. A related concern is agent scope: branch protection rules and access controls provide an important safeguard against agents taking destructive actions — such as merging to protected branches or modifying tracked artifacts outside their intended scope — and their configuration should be treated as a required component of any governance deployment.

As a validation case reaches a stable intermediate stage, condensing the agent's context and explicitly prompting it to critically review the current model and documentation was found to be a valuable practice: it helps the agent clean up loose ends, identify inconsistencies, and establish a coherent base — both a model base and a context base — before continuing development. Unrelated or ongoing work in the repository may otherwise be inadvertently picked up and overanalyzed by the agent, so it is important to maintain a clean and well-scoped working environment throughout the development process.

\subsection{Provenance, authorship, and log management}
\label{sec:4_provenance}

A principle that emerged clearly from this work is that the human directing the development activity is the author of the resulting contribution, regardless of what fraction of the code is mechanically produced by an AI agent. The person driving the change is responsible for the changes submitted for review. Under PLN-4005 Section~9.4.1, the contributor is responsible for change request content and the independent reviewer controls its disposition — an accountability structure this framing preserves directly. Simultaneously, NQA-1 SP~2.7-402.1 requires that verification be performed by an individual other than the originator. The human-as-author principle provides exactly that definition and preserves the accountability chain on which independent verification depends.

Recording this provenance at the commit level through human-readable session logs linked to the corresponding development issue ensures that the development history is stochastically reconstructable from repository artifacts, without a reliance on external documentation. In practice, however, we observed that AI-generated human-readable summaries of these logs abbreviated or obscured the agent's actual decision-making in notable ways. However, additional instructions can be placed into the \texttt{AGENTS.md} file to preserve critical decision-making points and other key prompts to create a more reliable audit record.

This traceability requirement also has practical implications for repository hygiene. Raw log files accumulate quickly and can impose meaningful storage overhead on both local and remote repositories without some limitation or retention policy. A practical approach is to preserve full logs for recent development activity while storing simplified or compressed versions of older logs, balancing long-term traceability against maintainability.

\subsection{Implications for SQA, NQA-1 compliance, future governance, and ecosystem integrity}
\label{sec:4_implications}
 
The work described herein suggests that existing SQA processes and NQA-1 SQA requirements, while developed under the assumption of human authorship,
are largely compatible with AI-assisted development when interpreted appropriately.
The core requirements — traceability, independent verification, documented procedures,
and accountability — map naturally onto the governance framework proposed here.
What this work clarifies is where those requirements must be enforced
structurally rather than procedurally, and where human judgment remains not just
preferable but irreplaceable.
 
One question that this work surfaced was whether AI coding
agents should themselves be treated as configuration items or support libraries within
the NQA-1 framework and be subject to the same qualification, versioning, and
change-control requirements that are applied to other software components that influence the
development process. As agentic tools become more deeply embedded in development
infrastructure, this question may continue to show up in standards bodies discussions and
project governance teams. However, it should be noted that current AI tools 
have stochastic output characteristics. Even with version-controlled inputs, there
are no assurances that you can reproduce the identical output even with the same
version of a given AI model. From this perspective, it is the opinion of these authors
that there is little added value in attempting to version control these models. However,
it could be useful to at least document the version used in a particular commit in
case post-analysis is useful in the future.
 
Beyond the immediate development cycle, the provenance framework developed here
also addresses a longer-term systemic risk. Research on model collapse has
demonstrated that the quality of generative AI systems trained on AI-generated outputs rather than
human-generated data progressively degrades~\cite{Shumailov2024,Gibney2024}.
In code generation, the analogous failure mode is the gradual amplification of
systematic error patterns as described in \cref{sec:4_hallucination}. These defaults
manifest as subtly defective AI-generated code enters
widely indexed open-source repositories and is subsequently scraped into future
training corpora. This risk is particularly acute in scientific software, where
domain-specific idioms and numerical implementation patterns are less
well-represented in general training data, which makes it harder for future models to
distinguish correct from plausible-but-wrong code. By ensuring that AI-generated
contributions are explicitly labeled and traceable within the version control history,
the framework proposed here enables downstream dataset curators to track provenance
and filter synthetic contributions where appropriate. Traceability is therefore not only
a present-day quality assurance requirement, but a contribution to the long-term
integrity of the scientific software ecosystem.
 
Taken together, the lessons documented in this section do not undermine the
framework proposed in \cref{sec:2_methodology}, they refine and extend it. The core
premise holds: V\&V cases, with their objective correctness criteria and bounded scope,
provide an effective proving ground for establishing and iterating on AI governance
practices that can later extend to broader scientific software development. The
framework's emphasis on traceability, human accountability, and structured constraints
on agent behavior was validated in practice. What this experience adds is a clearer
picture of where those constraints must be enforced at the infrastructure level, how
the developer's role evolves under agentic workflows, and why the governance of
AI-assisted scientific software development is a concern not only for the quality of
today's code, but for the integrity of the AI systems that will assist in writing
tomorrow's.

\section{Conclusion}
\label{sec:conclusion}

The transition to AI-assisted scientific software development is not approaching — 
it has arrived. Researchers and engineers are already using LLMs 
to write code, generate tests, and draft documentation, often without formal 
acknowledgment or governance. In safety-critical domains governed by rigorous SQA standards (e.g., NQA-1), where traceability, independent verification, and documented 
procedures are foundational requirements, this ad hoc practice represents a 
systemic and unacceptable risk.

This paper proposed a governance framework for AI-assisted development 
in NQA-1-governed (or any high SQA-conscious) open-source scientific software and demonstrated it through 
the development of two new validation cases for TMAP8, an advanced tritium 
transport code for fusion energy. The framework encodes disclosure, provenance, 
and review requirements in a lightweight repository-level \texttt{AGENTS.md} 
specification, enforces non-negotiable requirements through automated pre-commit 
hooks, and preserves human accountability throughout. By embedding these 
requirements directly in the development infrastructure rather than relying on 
developer discretion, the framework ensures that AI-assisted contributions are 
subject to the same standards as human-authored code — and that the degree 
of AI involvement is visible, auditable, and linked to the version control history.

The two validation (i.e., model predictions compared against experimental measurements) cases demonstrated both the practical productivity gains 
and the characteristic failure modes of governed AI-assisted development. In 
the first case, for which a published model existed, the agent significantly 
accelerated the mechanical aspects of implementation and Bayesian parameter 
optimization, while also reproducing an omission from the source publication 
that was only caught through adversarial human review. In the second case, 
for which no published model existed, the agent enabled rapid prototyping of 
competing physical formulations, expanding the hypothesis space explored 
within a realistic development timeline — a qualitatively different and 
particularly valuable form of AI assistance for exploratory scientific work. 
In both cases, structured guidance in \texttt{AGENTS.md} reduced hallucination 
frequency, while thorough human review remained indispensable for ensuring 
correctness, interpretability, lack of unnecessary complexity, and scientific fidelity.

Several cross-cutting lessons emerged from this experience. Hallucination is 
a persistent feature of AI-assisted workflows that does not disappear with 
better prompting alone; it requires structural countermeasures, including 
well-scoped task definitions, automated quality gates, and human reviewers 
capable of reasoning beyond the source material the agent was given. Context 
limitations cause agents to overlook explicitly documented requirements, 
making infrastructure-level enforcement — through pre-commit hooks rather 
than agent instructions alone — more reliable for non-optional provenance 
requirements. The developer's role shifts meaningfully under agentic workflows, 
from direct code authorship toward prompt engineering, output review, and 
active scientific oversight; the cognitive savings on individual tasks do not 
reduce the overall burden of maintaining quality. And the human directing 
the work remains the author and bears full responsibility for what is submitted 
for review, regardless of how much of the artifact was mechanically produced 
by an agent.

Looking ahead, the provenance framework developed here carries 
implications beyond the immediate development cycle. As AI-generated code 
enters public repositories and is scraped into future training corpora, 
the risk of unreviewed or mislabeled synthetic contributions degrading the quality 
of the very models that future developers will rely on increases. Explicit, 
machine-readable provenance records are therefore not only a present-day 
quality assurance requirement, but a contribution to the long-term integrity 
of the scientific software ecosystem.

V\&V cases, with their objective correctness criteria, bounded scope, and 
full auditability, proved to be the right proving ground for establishing 
these governance practices. The same properties that make V\&V cases effective 
for catching human coding errors — known reference solutions, measurable 
tolerances, and structured documentation requirements — make them 
well suited for evaluating and governing AI-generated code.

The relevant choice facing the scientific software community is not between 
AI-assisted and human-only development. It is between governed and ungoverned 
AI assistance. The framework presented here demonstrates that 
AI-assisted scientific software development can be responsibly governed today within existing quality assurance standards, and the productivity gains that motivate AI adoption in the first place do not have to be sacrificed in the process. The framework 
proposed here is designed to be iterative: \texttt{AGENTS.md} is a living 
specification, updated as new failure modes are identified, and the 
governance practices it encodes are expected to evolve alongside the 
agentic tools they govern and our understanding of their advantages and risks. 

\section*{Data availability}

The TMAP8 code is open source and available at \url{https://github.com/idaholab/TMAP8}. It contains the \texttt{AGENTS.md} file and all the input files and python scripts used to generate the simulation results and figures in this manuscript \cite{tmap8-pr-397,tmap8-pr-402,tmap8-pr-409}. The complete TMAP8 documentation, including instructions on how to get started, the validation cases cases described above, and descriptions of TMAP8's capabilities, is available at \url{https://mooseframework.inl.gov/TMAP8/index.html}.

\section*{Acknowledgements}

This work was supported through INL’s Laboratory Directed Research \& Development (LDRD) Program under DOE Idaho Operations Office Contract DE-AC07-05ID14517. The United States Government retains, and the publisher, by accepting the article for publication, acknowledges that the United States Government retains a nonexclusive, paid-up, irrevocable, worldwide license to publish or reproduce the published form of this manuscript, or allow others to do so, for United States Government purposes.

This research made use of Idaho National Laboratory's High Performance Computing systems located at the Collaborative Computing Center and supported by the Office of Nuclear Energy of the U.S. Department of Energy and the Nuclear Science User Facilities under Contract No. DE-AC07-05ID14517.

\bibliography{references}

\section*{Supplementary material}

\subsection*{Validation case 1: Tritium release from neutron-irradiated Li$_2$TiO$_3$}
\label{sec:validation:1}

This validation case models TDS from neutron-irradiated Li$_2$TiO$_3$ (lithium titanate) crystalline grains, a candidate solid tritium breeding material for D-T fusion energy systems since lithium generates tritium when hit by fusion neutrons. The Li$_2$TiO$_3$ samples were irradiated at the Kyoto University Reactor at various neutron fluences \cite{kobayashi2015developing}. After irradiation, the tritium release behavior was measured by TDS with a heating rate of \SI{5}{K/min} starting from \SI{300}{K} using pure helium as the purge gas. The average grain radius was \SI{1.5}{\micro m}. Irradiation has a significant effect on tritium transport. Tritium atoms tend to interact with defects in their host materials. These defects (e.g., interstitial sites, vacancies, dislocations, grain boundaries, pores) can trap tritium atoms and slow down tritium transport. By creating defects in materials, irradiation can significantly affect tritium transport in solid materials. 
Sample~E from Kobayashi et al. \cite{kobayashi2015developing}, which contains a high defect density ($D_{id} = \SI{3.384e26}{at/m^3}$), is modeled. Tritium release is significantly influenced by trapping at O$^{-}$-centers and defect annihilation at higher temperature.

\subsubsection*{Model description} \label{sec:validation:1:model_description}

In this section, we present the governing equations for tritium transport in a spherical sample, including mobile tritium diffusion, trapping defect annihilation, trapping and detrapping, as well as the boundary and initial conditions and model parameters.

\paragraph{Diffusion of mobile species}

TMAP8 simulates tritium diffusion and trapping in a single spherical grain of Li$_2$TiO$_3$ using one-dimensional spherical coordinates. The governing equation for mobile tritium concentration~$C$ is~\cite{kobayashi2015developing}:
\begin{equation} \label{eq:val1:diffusion_trapping}
\frac{\partial C}{\partial t} = D \left( \frac{\partial^2 C}{\partial r^2} + \frac{2}{r} \frac{\partial C}{\partial r} \right) - \frac{\partial C_T}{\partial t},
\end{equation}
where $C_T$ is the concentration of trapped tritium in O$^{-}$-centers and $D$ is the temperature-dependent diffusivity following the Arrhenius law:
\begin{equation} \label{eq:val1:diffusivity}
D = D_0 \exp \left( -\frac{E_d}{k_B T} \right),
\end{equation}
where $D_0$ is the pre-exponential factor, $E_d$ is the activation energy, $k_B$ is the Boltzmann constant, and $T$ is the temperature.

\paragraph{Defect annihilation}

During TDS heating, radiation-induced defect sites undergo first-order annihilation~\cite{kobayashi2015developing}:
\begin{equation} \label{eq:val1:annihilation}
\frac{d D_{id}}{dt} = -k_{dp\text{-}da} \, D_{id},
\end{equation}
where $D_{id}$ is the defect density and $k_{dp\text{-}da}$ is the annihilation rate coefficient. The trap site fraction~$\chi$ is related to~$D_{id}$; however, the exact relationship is not clearly indicated in~\cite{kobayashi2015developing}. Therefore, the initial trap site density is assumed to equal the defect density, i.e., $\chi(0)N = D_{id}$, where $N$ is the lattice site density.

The annihilation rate coefficient is described as:
\begin{equation} \label{eq:val1:annihilation_rate}
k_{dp\text{-}da} = k_{dp\text{-}da,0} \exp \left( -\frac{E_{dp\text{-}da}}{k_B T} \right),
\end{equation}
where $k_{dp\text{-}da,0}$ and $E_{dp\text{-}da}$ are the annihilation prefactor and energy, respectively. Raising the temperature reduces the available trap sites: the trap site fraction~$\chi$ decays over time following \cref{eq:val1:annihilation}, preventing retrapping into annihilated sites.

\paragraph{Trapping and detrapping}

Only O$^{-}$-center (hydroxyl group) trapping is included in this model. As noted by Kobayashi et al.~\cite{kobayashi2015developing}, tritium release controlled by detrapping from F$^+$-centers (oxygen vacancies) occurs near \SI{580}{K}, which corresponds to the release temperature controlled by the diffusion process itself. Because F$^+$-center detrapping is not rate-limiting relative to diffusion, it does not produce a distinct feature (i.e., peak) in the TDS spectrum and is therefore excluded from the model.

The trapped concentration~$C_T$ evolves according to:
\begin{equation} \label{eq:val1:trapping}
\frac{\partial C_T}{\partial t} = \alpha_t \frac{C_T^{\text{empty}} \, C}{N} - \alpha_r \, C_T - k_{dp\text{-}da} \, C_T,
\end{equation}
where $C_T^{\text{empty}} = \chi N - C_T$ is the empty trap concentration. The trapping and detrapping rate coefficients follow Arrhenius relationships:
\begin{equation} \label{eq:val1:trapping_rate}
\alpha_t = \alpha_{t0} \exp\left(-\frac{\epsilon_t}{k_B T}\right),
\end{equation}
\begin{equation} \label{eq:val1:detrapping_rate}
\alpha_r = \alpha_{r0} \exp\left(-\frac{\epsilon_r}{k_B T}\right),
\end{equation}
where $\alpha_{t0}$ and $\alpha_{r0}$ are the pre-exponential factors for the trapping and release rate coefficients, and $\epsilon_t$ and $\epsilon_r$ are the trapping and release energies. The last term in \cref{eq:val1:trapping}, $k_{dp\text{-}da} \, C_T$, accounts for the release of trapped tritium when defect sites are annealed.

\paragraph{Boundary and initial conditions}

The boundary conditions are $\partial C / \partial r = 0$ at $r = 0$ due to the symmetry at grain center and $C = 0$ at $r = r_g$ due to the fast surface release.
The mobile and trapped tritium concentrations are initialized at their local trapping/detrapping equilibrium values at the starting temperature $T_{\text{start}} = \SI{300}{K}$. The equilibrium mobile concentration is computed from the balance of trapping and detrapping rates, avoiding an initial transient from any imbalance. Since the TDS output is normalized to arbitrary units, only the shape of the release curve matters, not the absolute concentrations.

\paragraph{Case and model parameters} \label{sec:validation:1:parameters}

All model parameters are taken from Kobayashi et al.~\cite{kobayashi2015developing} and summarized in \cref{tab:val1:parameters}.

\begin{table}[htbp]
\centering
\caption{Values of model parameters for validation case~1.}
\label{tab:val1:parameters}
\begin{tabular}{llrll}
\toprule
Parameter & Description & Value & Units \\
\midrule
$r_g$ & Grain radius & 1.5 & \si{\micro m}\\
$D_0$ & Diffusivity pre-exponential & $6.9 \times 10^{-7}$ & \si{m^2/s}\\
$E_d$ & Diffusion activation energy & 1.07 & \si{eV}\\
$\alpha_{t0}$ & Trapping prefactor & $4.2 \times 10^{8}$ & \si{s^{-1}}\\
$\epsilon_t$ & Trapping energy & 1.04 & \si{eV}\\
$\alpha_{r0}$ & Detrapping prefactor & $4.1 \times 10^{6}$ & \si{s^{-1}}\\
$\epsilon_r$ & Detrapping energy & 1.19 & \si{eV}\\
$k_{dp\text{-}da,0}$ & Annihilation prefactor & $1.0 \times 10^{2}$ & \si{s^{-1}}\\
$E_{dp\text{-}da}$ & Annihilation energy & 0.9 & \si{eV}\\
$N$ & Lattice density & $1.88 \times 10^{28}$ & \si{at/m^3}\\
$\beta$ & TDS heating rate & 5 & \si{K/min}\\
$D_{id,E}$ & Defect density (Sample~E) & $3.384 \times 10^{26}$ & \si{at/m^3}\\
\bottomrule
\end{tabular}
\end{table}

\subsubsection*{Results} \label{sec:validation:1:results}

In this section, we discuss the TDS tritium release spectrum using both the initial parameters from Kobayashi et al.~\cite{kobayashi2015developing} and the parameters optimized by Bayesian optimization.

\paragraph{Results before optimization}

\begin{figure}[htbp]
    \centering
    \includegraphics[width=0.6\linewidth]{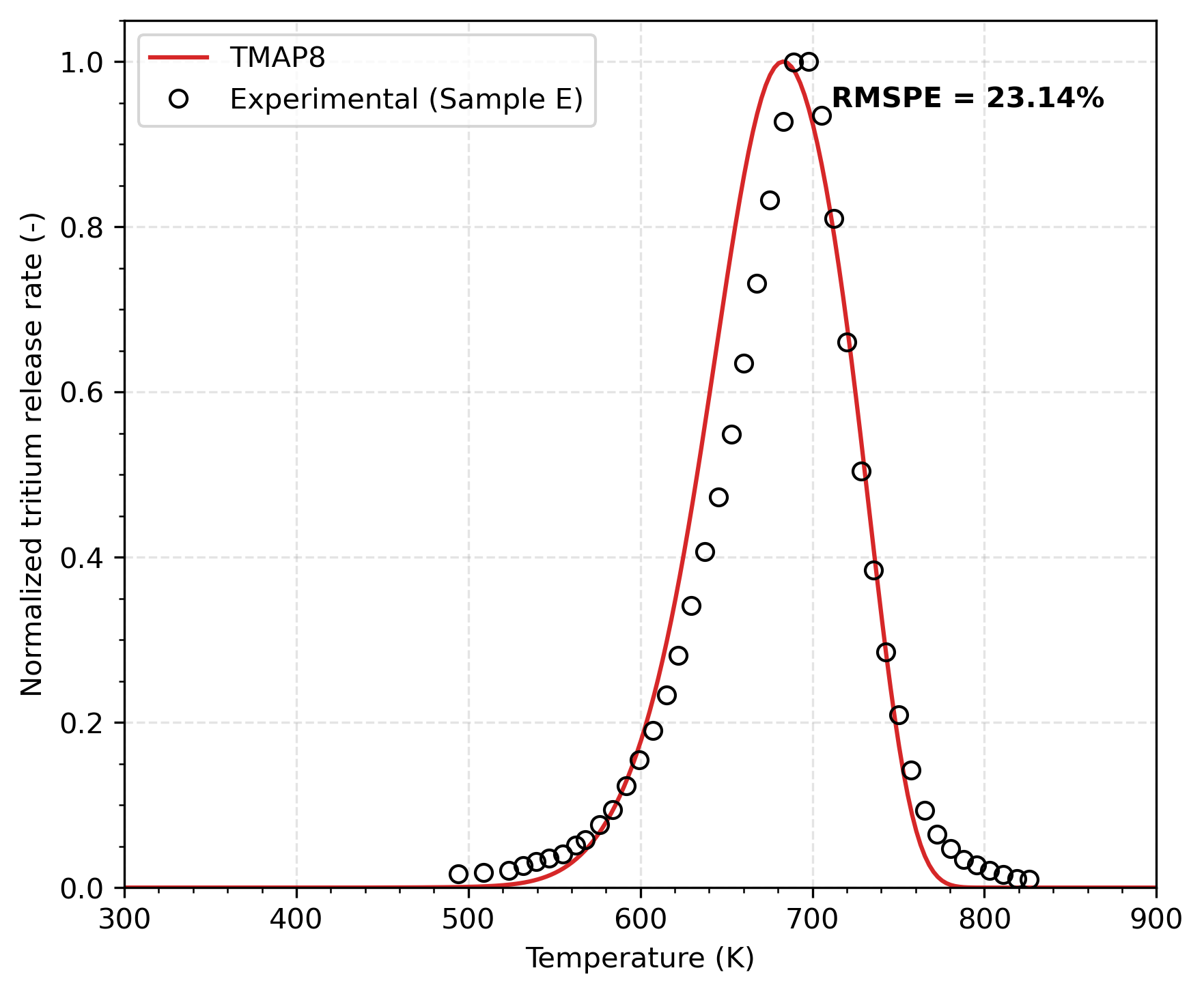}
    \caption{Comparison of TMAP8 calculation with the experimental TDS data for Sample~E (high defect density). Experimental data from Kobayashi et al.~\cite{kobayashi2015developing}.}
    \label{fig:val1:comparison_ref}
\end{figure}

\Cref{fig:val1:comparison_ref} compares TMAP8 and the experimental TDS spectrums for Sample~E. The O$^{-}$-center trapping model with defect annihilation captures the broad release profile. The high detrapping energy of O$^{-}$-centers (\SI{1.19}{eV}) produces a release peak above \SI{650}{K} that is distinct from the diffusion-controlled release. The defect annihilation mechanism reduces the effective trap density at high temperatures, suppressing retrapping and allowing tritium to escape more efficiently.

\paragraph{Results after optimization}

The agreement between the TMAP8 simulation and experimental data can be improved by optimizing the model parameters using the MOOSE stochastic tools module. A Bayesian optimization approach~\cite{DHULIPALA2026102776} was applied to optimize eight key parameters (i.e., four Arrhenius pre-exponential factors in $\log_{10}$ space and four activation energies for the diffusivity, trapping, releasing, and defect annealing) to better match the experimental TDS curve for Sample~E.

\begin{figure}[htbp]
    \centering
    \includegraphics[width=0.6\linewidth]{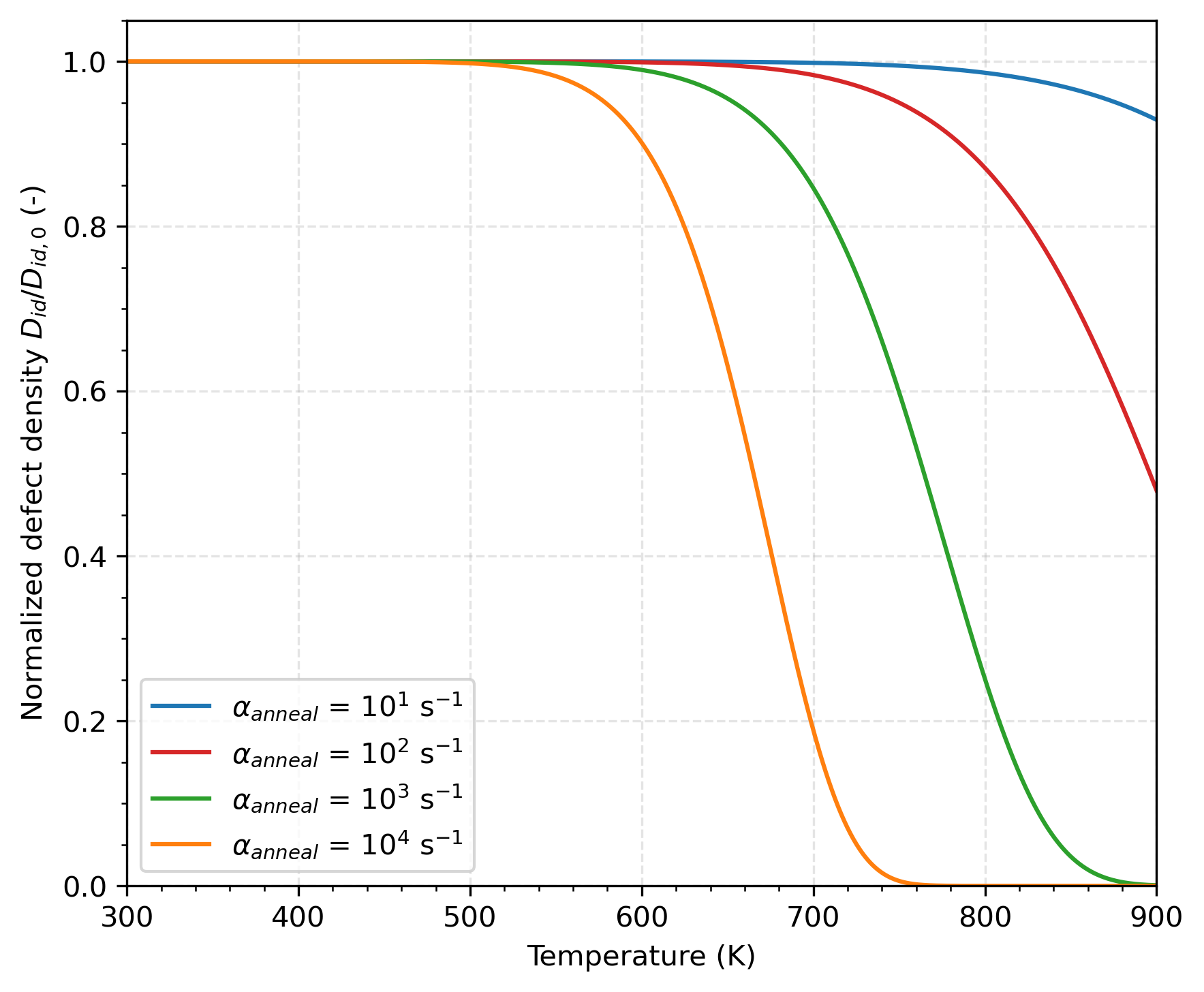}
    \caption{Evolution of the normalized defect density, $D_{id}/D_{id,0}$, during the TDS temperature ramp. As expected, the annihilation temperature strongly depends on $\alpha_{anneal}$.}
    \label{fig:val1:defect_density}
\end{figure}

As shown in \cref{fig:val1:defect_density}, the normalized defect density with the reference annihilation prefactor ($\alpha_{anneal} = \SI{e2}{s^{-1}}$) remains close to unity throughout the main release region (below ${\sim}\SI{750}{K}$) and only decreases significantly at higher temperatures where the tritium release flux is decreasing. Larger annihilation prefactors would shift the defect annihilation and the associated tritium release to lower temperatures, but the optimization consistently finds values near the reference.

The optimization used Gaussian process active learning with expected improvement acquisition, running 40 iterations with 5 parallel proposals per iteration. The objective function evaluates the root mean square percentage error (RMSPE) between the simulated and experimental normalized release rates using a continuous comparison at every simulation timestep. The experimental TDS curve is represented as a piecewise-linear interpolation function, and the RMSPE is accumulated over the full temperature ramp. Low-temperature constraint points (\SIrange{300}{475}{K}) with a small target value penalize parameter sets that produce spurious early release peaks.

\Cref{tab:val1:optimized_parameters} compares the reference values from Kobayashi et al.~\cite{kobayashi2015developing} with the Bayesian-optimized parameters, along with the parameter ranges used in the optimization. These parameters include the diffusivity prefactor ($D_0$) and activation energy ($E_d$), trapping rate prefactor ($\alpha_{t0}$) and activation energy ($\epsilon_t$), detrapping rate prefactor ($\epsilon_{r0}$) and activation energy ($\epsilon_r$), and annihilation rate prefactor ($k_{dp\text{-}da,0}$) and activation energy ($E_{dp\text{-}da}$).

\begin{table}[htbp]
\centering
\caption{Reference and Bayesian-optimized parameter values for validation case~1.}
\label{tab:val1:optimized_parameters}
\begin{tabular}{lrrrl}
\toprule
Parameter & Reference & Optimized & Range & Units \\
\midrule
$D_0$ & $6.9 \times 10^{-7}$ & $4.50 \times 10^{-6}$ & $10^{-8}$--$10^{-4}$ & \si{m^2/s} \\
$E_d$ & 1.07 & 1.01 & 0.8--1.4 & \si{eV} \\
$\alpha_{t0}$ & $4.2 \times 10^{8}$ & $2.21 \times 10^{7}$ & $10^{7}$--$10^{10}$ & \si{s^{-1}} \\
$\epsilon_t$ & 1.04 & 0.82 & 0.8--1.3 & \si{eV} \\
$\alpha_{r0}$ & $4.1 \times 10^{6}$ & $2.14 \times 10^{5}$ & $10^{5}$--$10^{8}$ & \si{s^{-1}} \\
$\epsilon_r$ & 1.19 & 1.08 & 0.9--1.5 & \si{eV} \\
$k_{dp\text{-}da,0}$ & $1.0 \times 10^{2}$ & $8.26 \times 10^{1}$ & $10^{0}$--$10^{5}$ & \si{s^{-1}} \\
$E_{dp\text{-}da}$ & 0.9 & 1.27 & 0.5--1.5 & \si{eV} \\
\bottomrule
\end{tabular}
\end{table}

\begin{figure}[htbp]
    \centering
    \begin{subfigure}[b]{0.35\linewidth}
        \includegraphics[width=\linewidth]{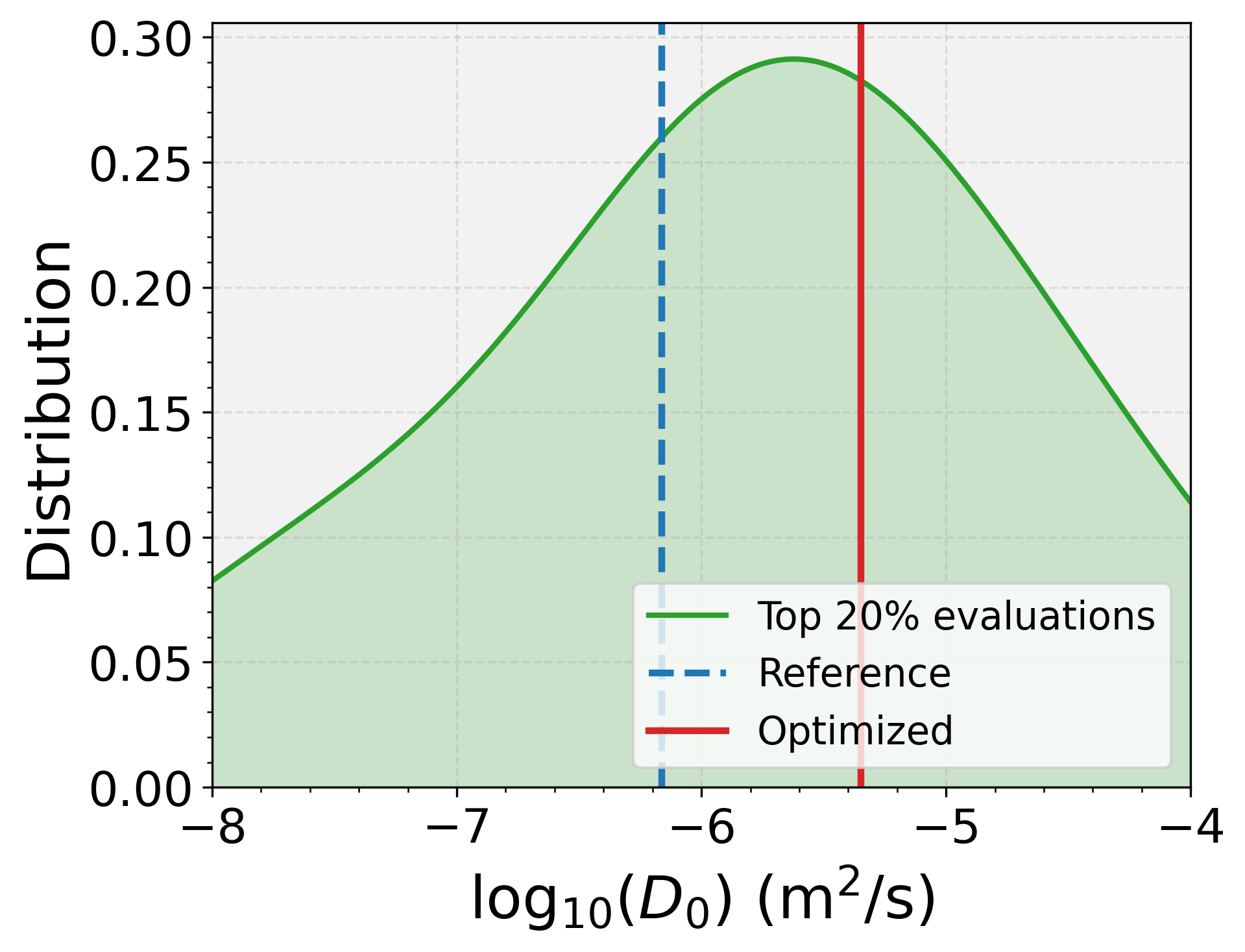}
        \caption{$D_0$}
    \end{subfigure}
    \begin{subfigure}[b]{0.35\linewidth}
        \includegraphics[width=\linewidth]{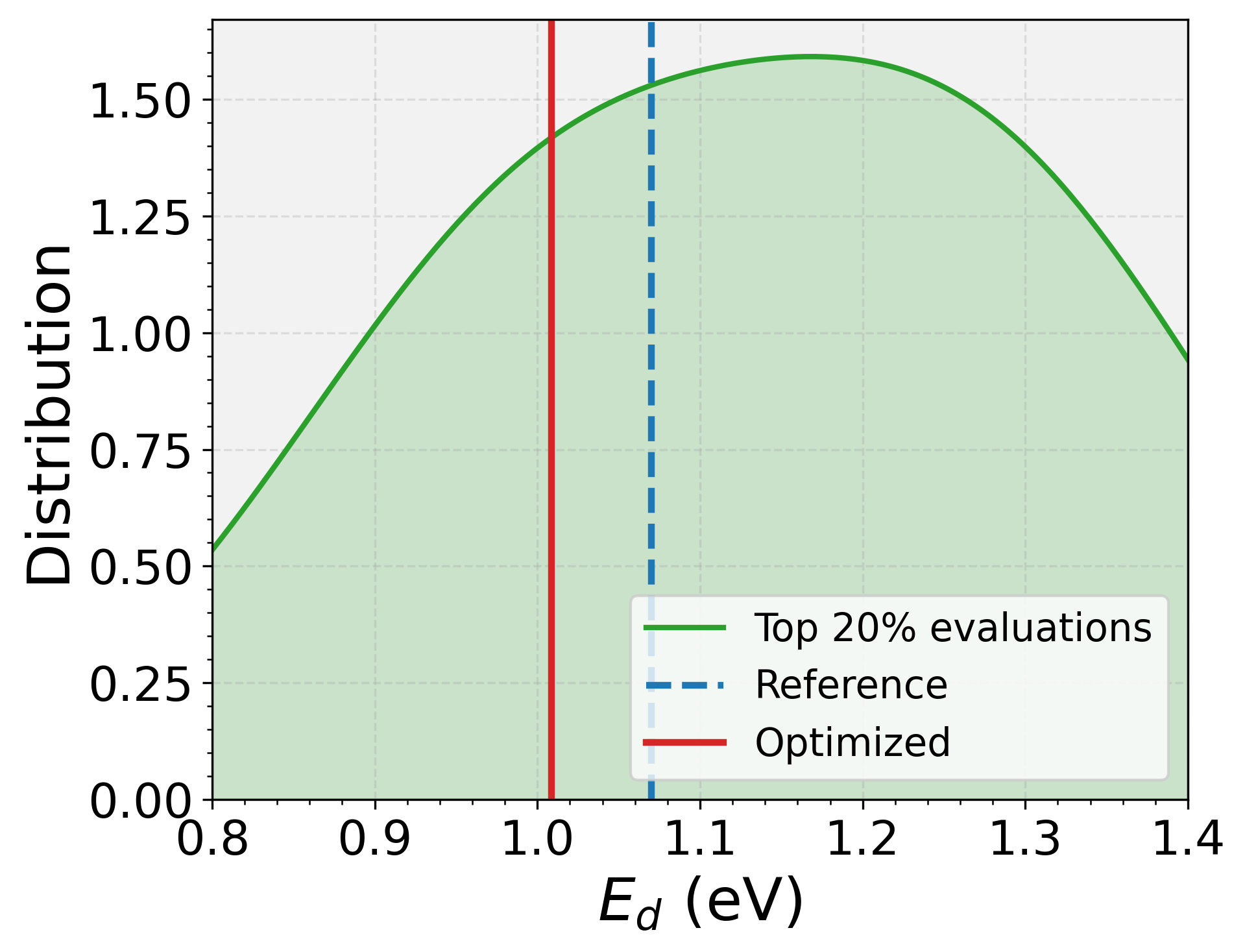}
        \caption{$E_d$}
    \end{subfigure}\\
    \begin{subfigure}[b]{0.35\linewidth}
        \includegraphics[width=\linewidth]{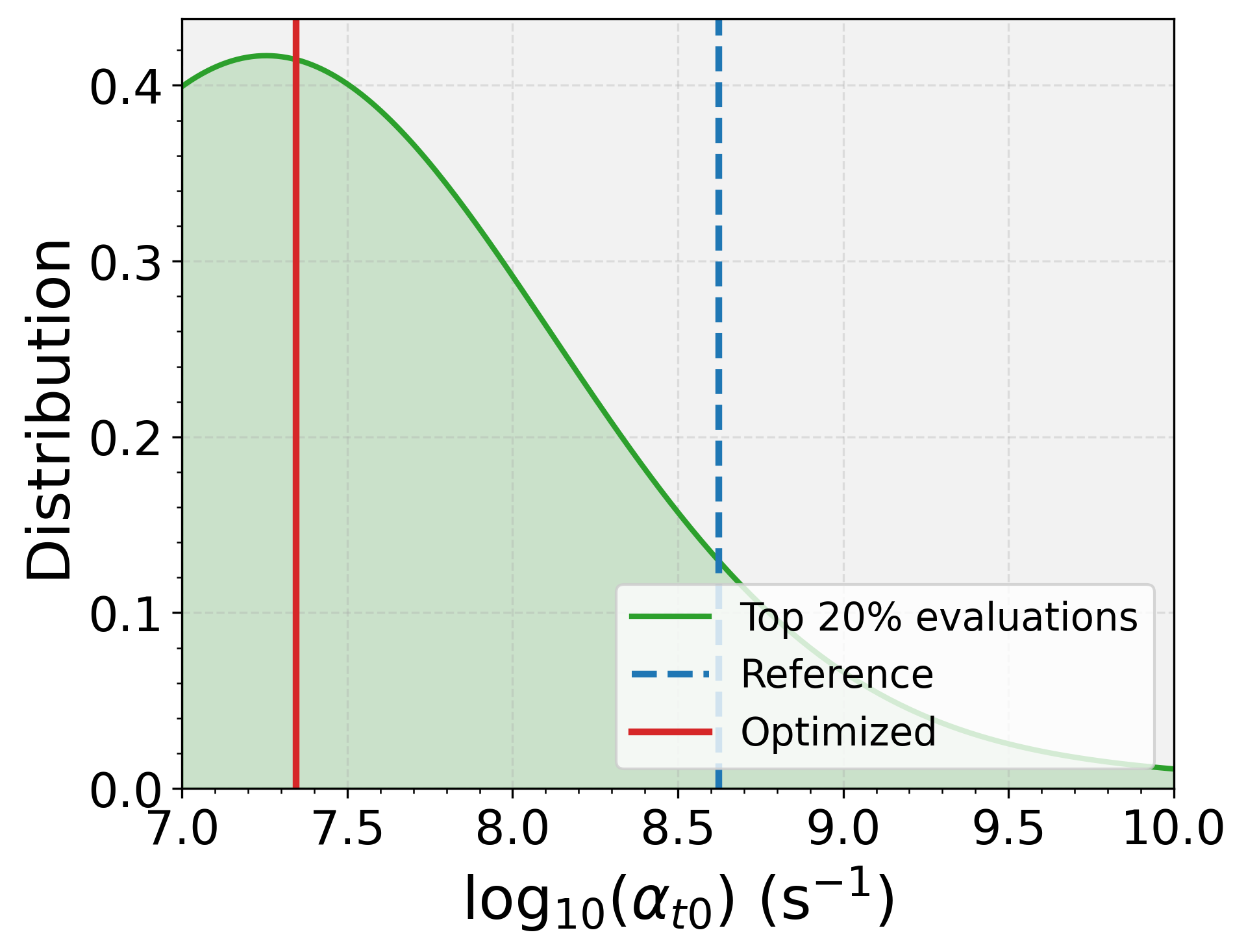}
        \caption{$\alpha_{t0}$}
    \end{subfigure}
    \begin{subfigure}[b]{0.35\linewidth}
        \includegraphics[width=\linewidth]{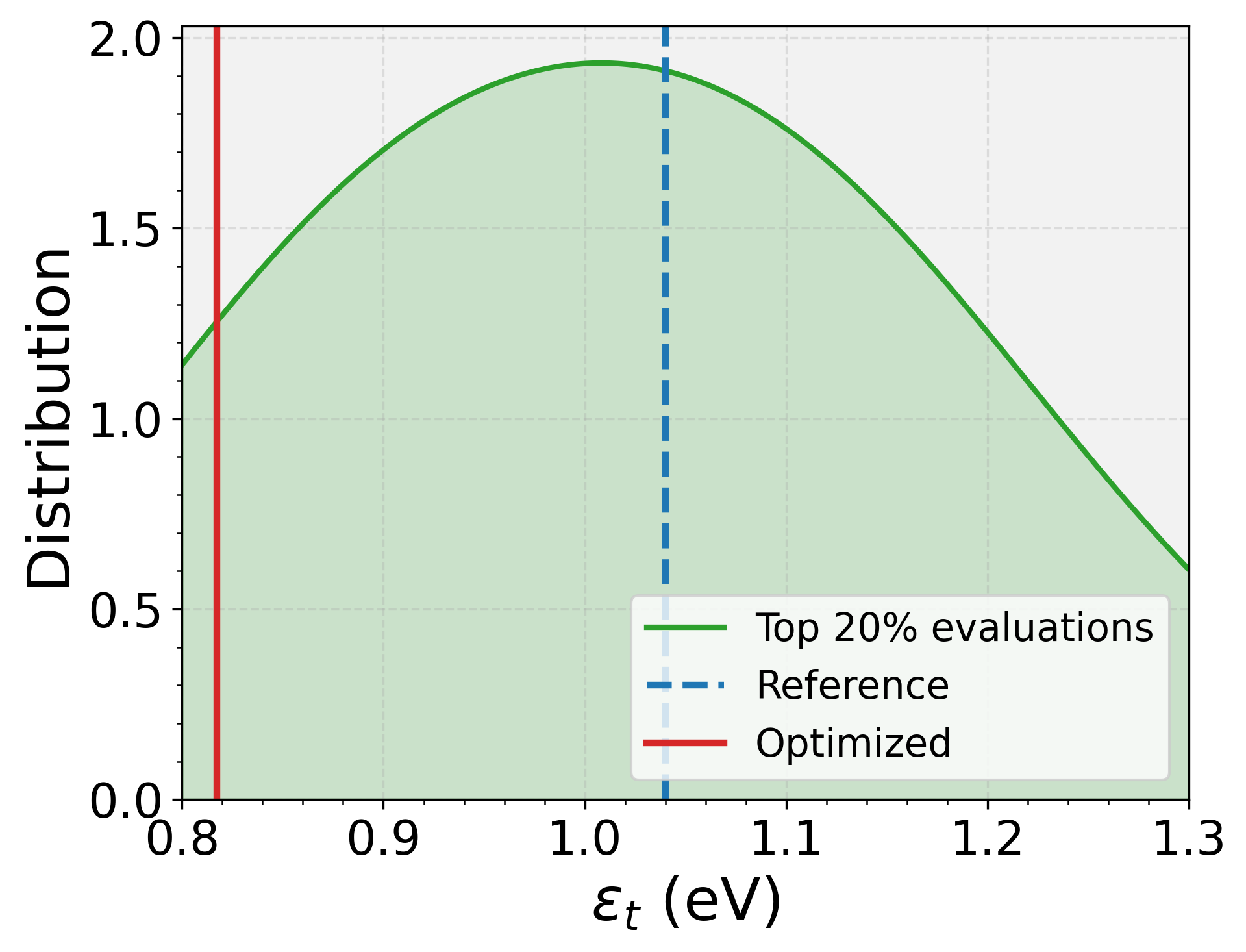}
        \caption{$\epsilon_t$}
    \end{subfigure}\\
    \begin{subfigure}[b]{0.35\linewidth}
        \includegraphics[width=\linewidth]{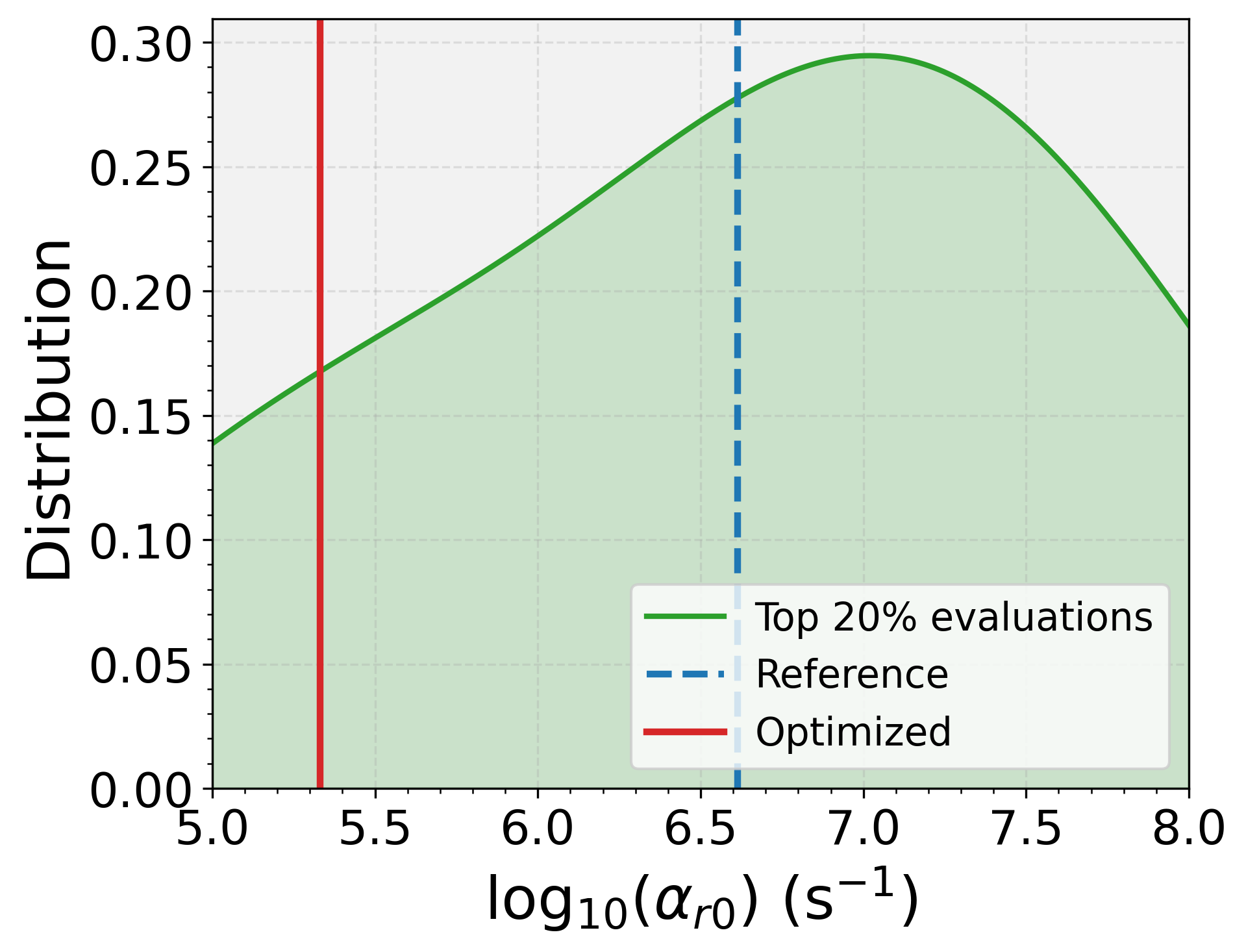}
        \caption{$\alpha_{r0}$}
    \end{subfigure}
    \begin{subfigure}[b]{0.35\linewidth}
        \includegraphics[width=\linewidth]{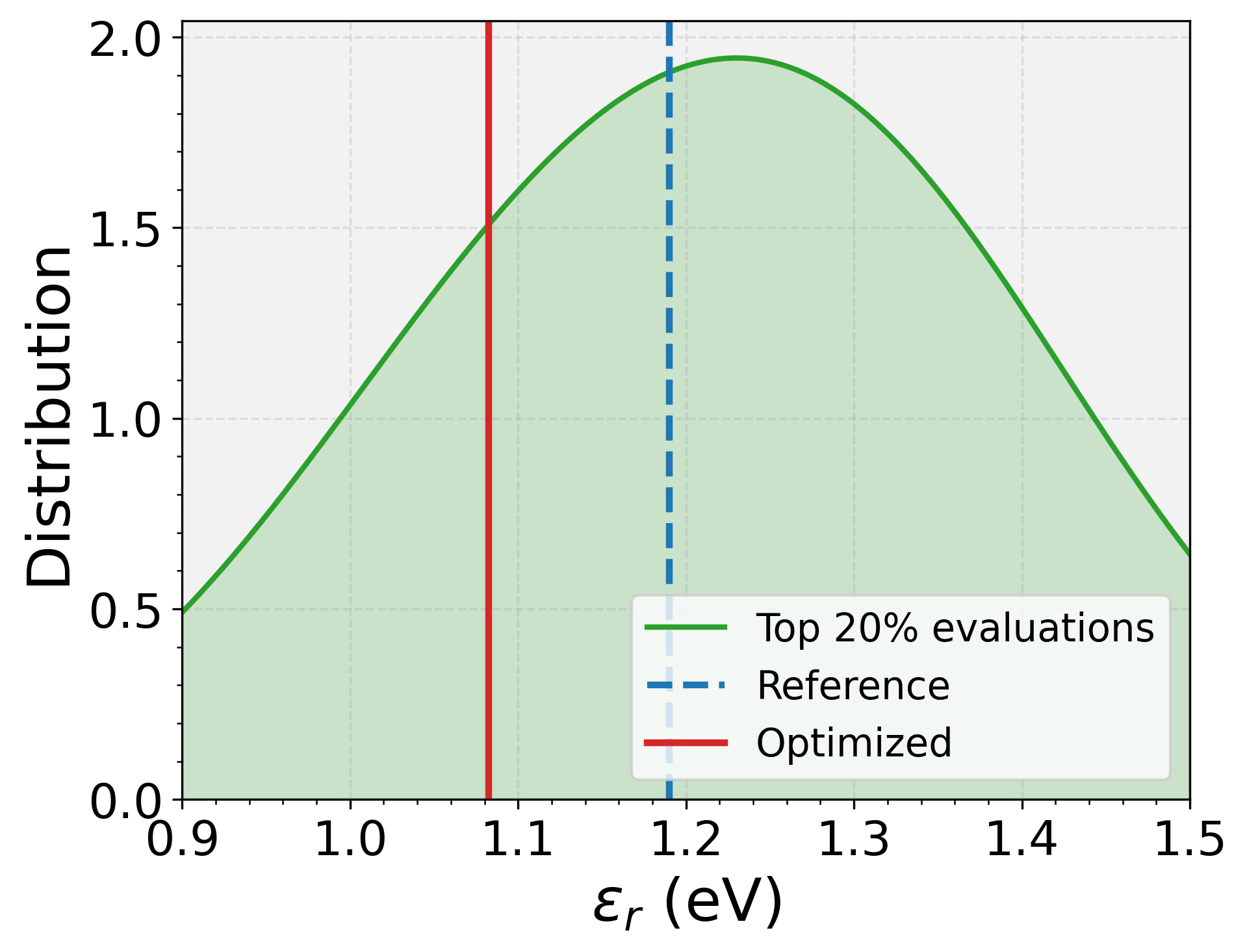}
        \caption{$\epsilon_r$}
    \end{subfigure}\\
    \begin{subfigure}[b]{0.35\linewidth}
        \includegraphics[width=\linewidth]{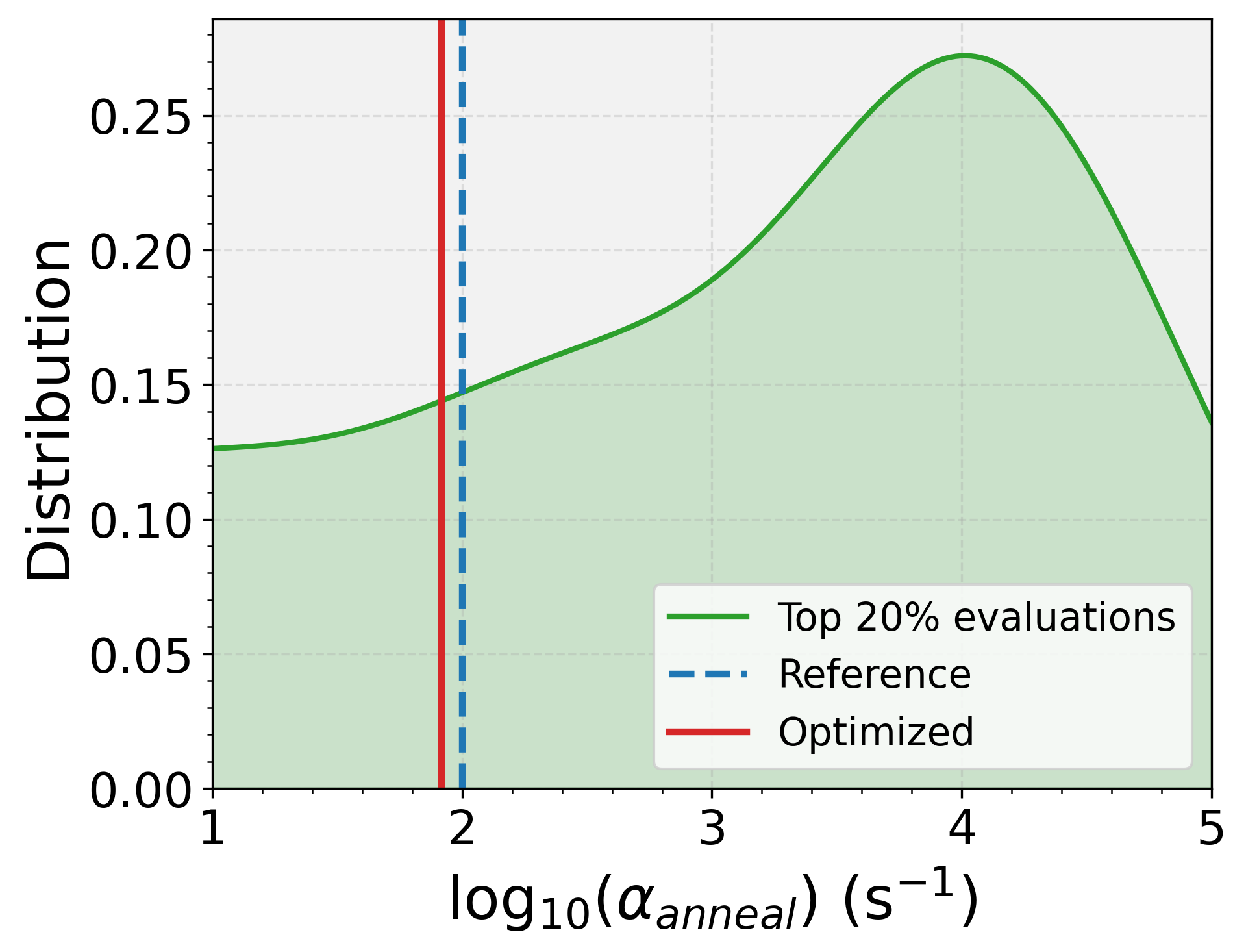}
        \caption{$k_{dp\text{-}da,0}$}
    \end{subfigure}
    \begin{subfigure}[b]{0.35\linewidth}
        \includegraphics[width=\linewidth]{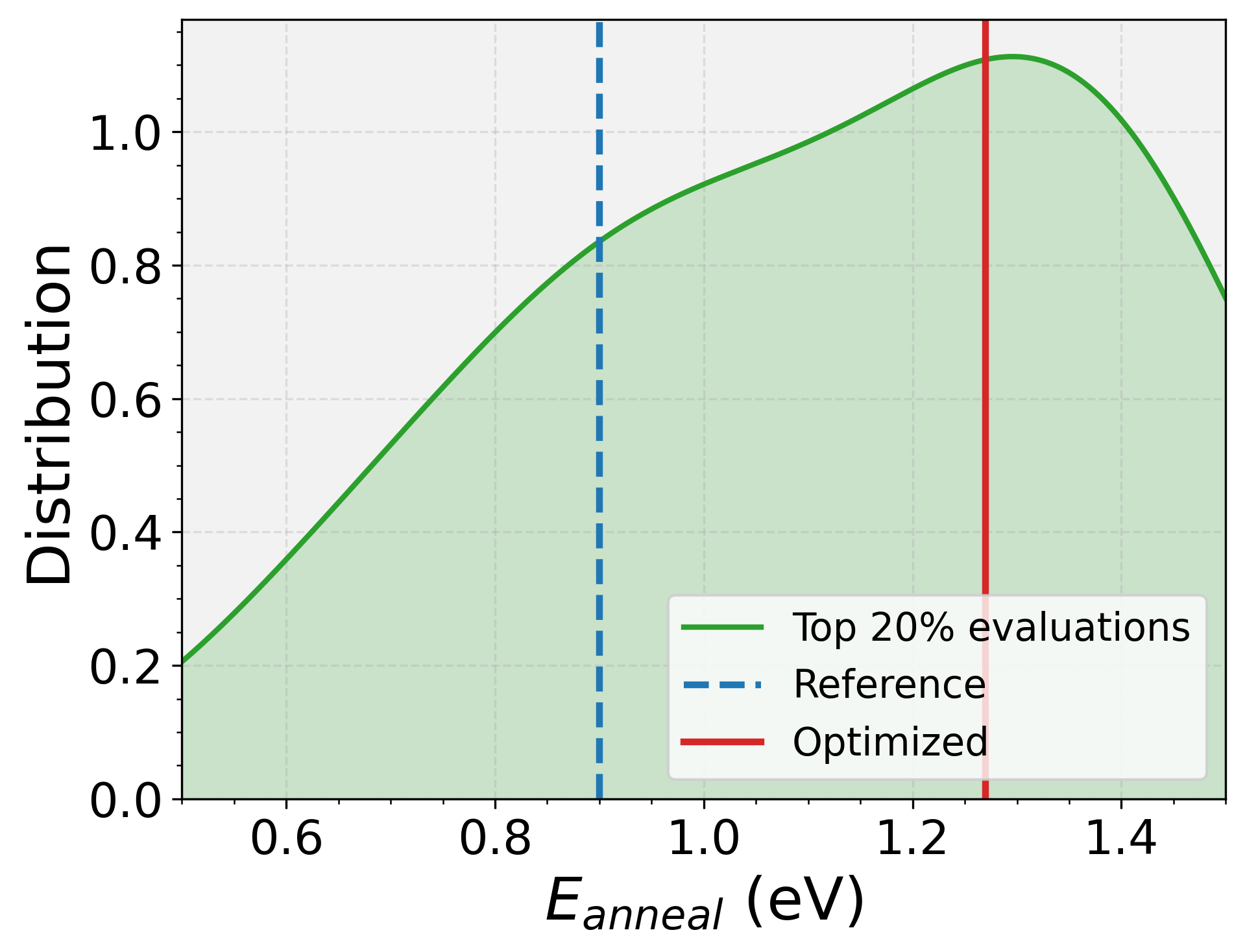}
        \caption{$E_{dp\text{-}da}$}
    \end{subfigure}
    \caption{Comparison of reference (blue dashed) and Bayesian-optimized (red solid) parameter values for (a)~diffusivity pre-exponential~$D_0$, (b)~diffusion activation energy~$E_d$, (c)~trapping prefactor~$\alpha_{t0}$, (d)~trapping energy~$\epsilon_t$, (e)~detrapping prefactor~$\alpha_{r0}$, (f)~detrapping energy~$\epsilon_r$, (g)~annihilation prefactor~$k_{dp\text{-}da,0}$, and (h)~annihilation energy~$E_{dp\text{-}da}$. Green curves show the kernel density estimate of the top 20\% scoring evaluations. The gray shaded region indicates the search range.}
    \label{fig:val1:param_exploration}
\end{figure}

\Cref{fig:val1:param_exploration} compares the reference parameter values from Kobayashi et al.~\cite{kobayashi2015developing} (blue dashed lines) with the Bayesian-optimized values (red solid lines) for each of the eight fitted parameters. The green curves show the kernel density estimate of the top 20\% scoring evaluations from the Bayesian optimization, providing insight into which parameter regions produce good fits to the experimental data. The optimized values fall within the high-density regions of the distributions, confirming consistency with the near-optimal parameter space.

\begin{figure}[htbp]
    \centering
    \begin{subfigure}[b]{0.48\linewidth}
        \includegraphics[width=\linewidth]{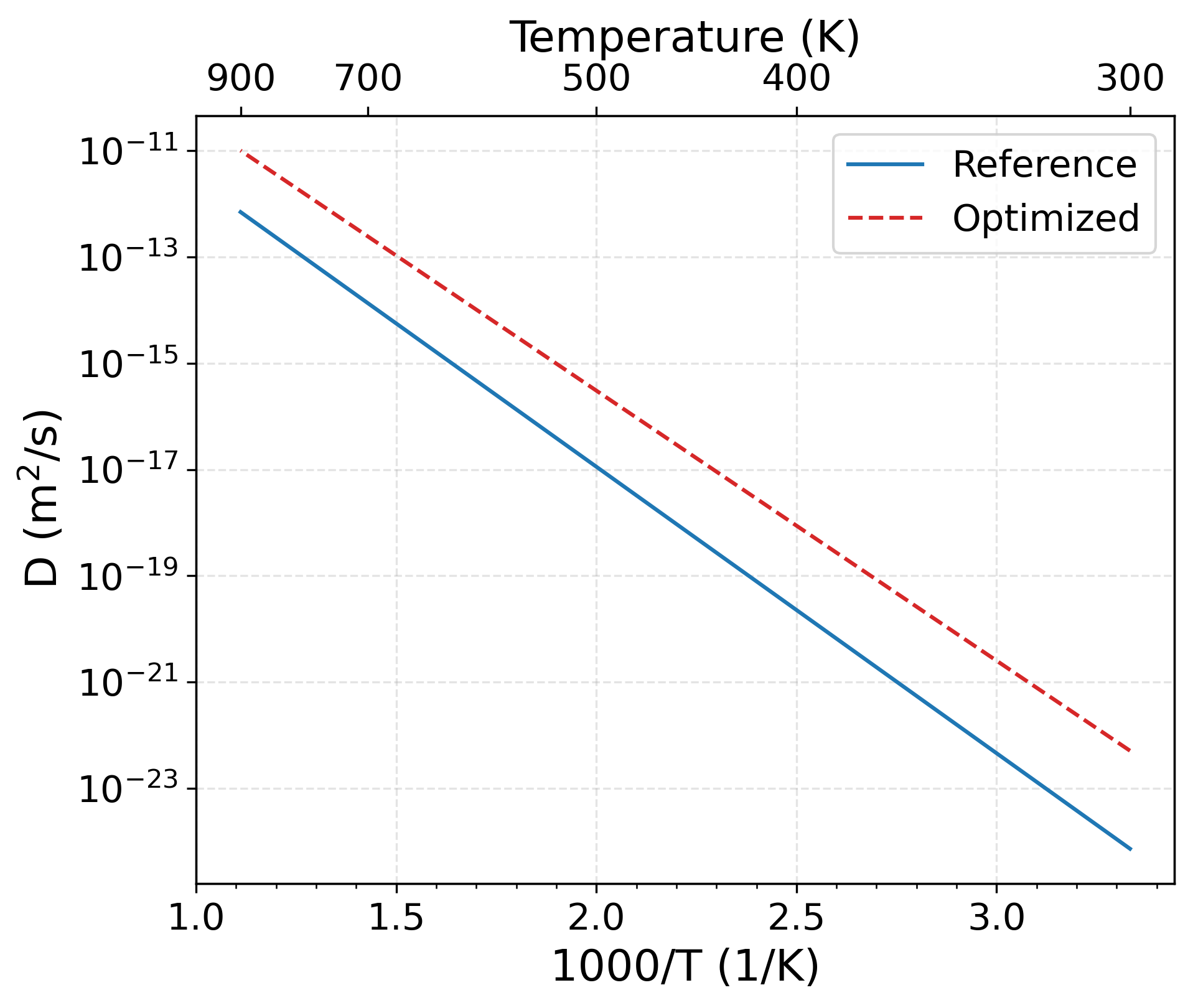}
        \caption{$D(T)$}
    \end{subfigure}%
    \hfill
    \begin{subfigure}[b]{0.48\linewidth}
        \includegraphics[width=\linewidth]{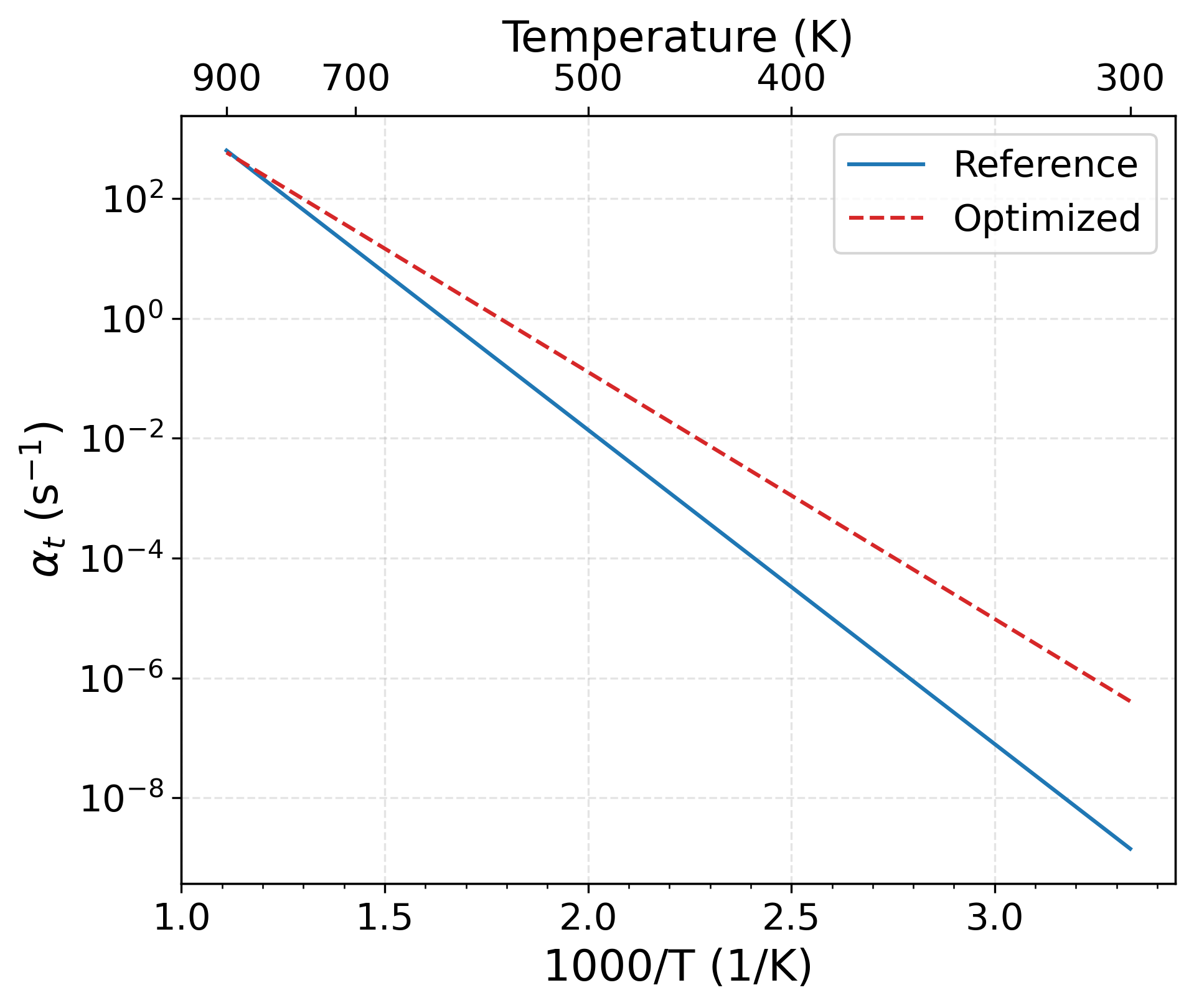}
        \caption{$\alpha_t(T)$}
    \end{subfigure}\\
    \begin{subfigure}[b]{0.48\linewidth}
        \includegraphics[width=\linewidth]{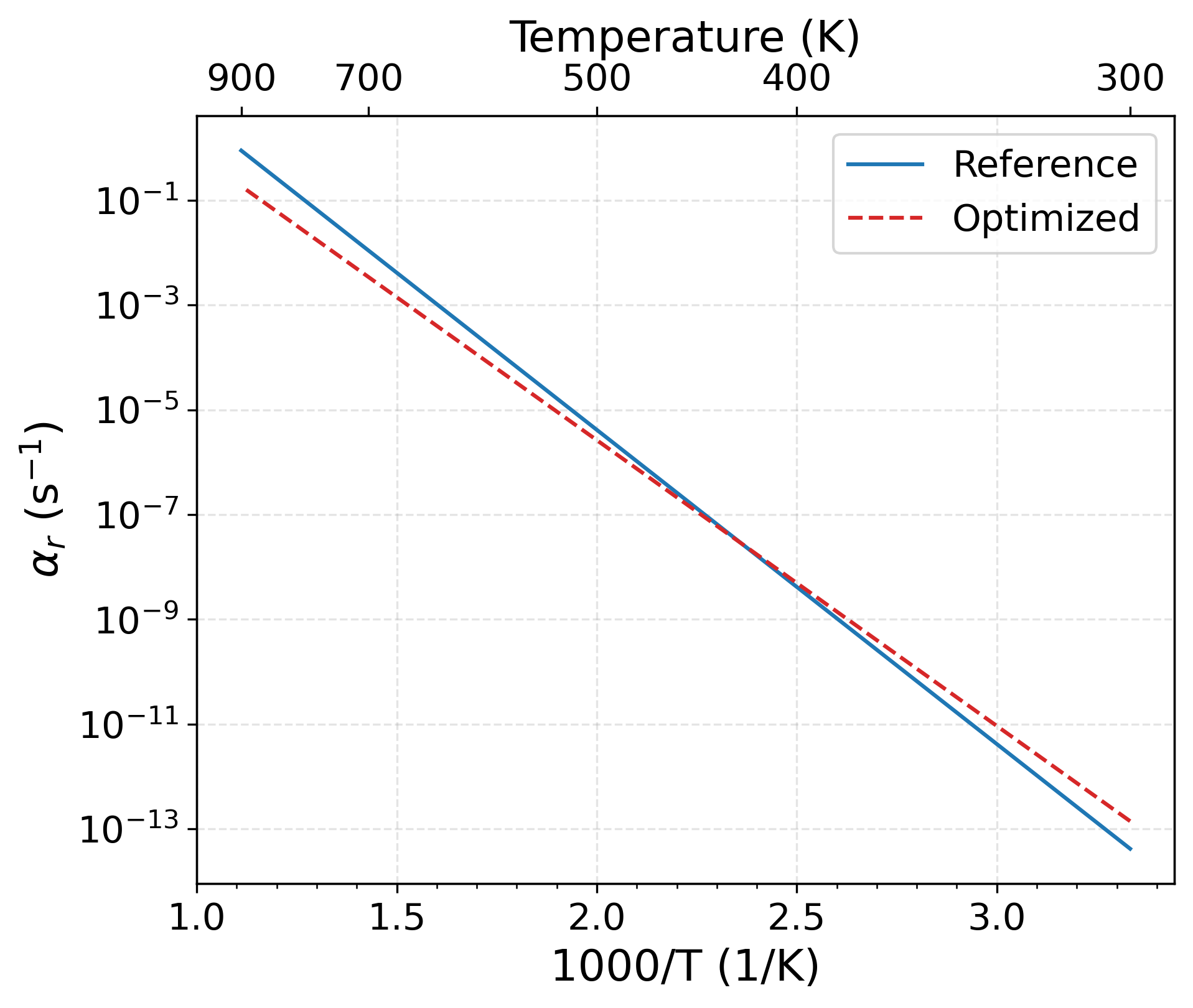}
        \caption{$\alpha_r(T)$}
    \end{subfigure}%
    \hfill
    \begin{subfigure}[b]{0.48\linewidth}
        \includegraphics[width=\linewidth]{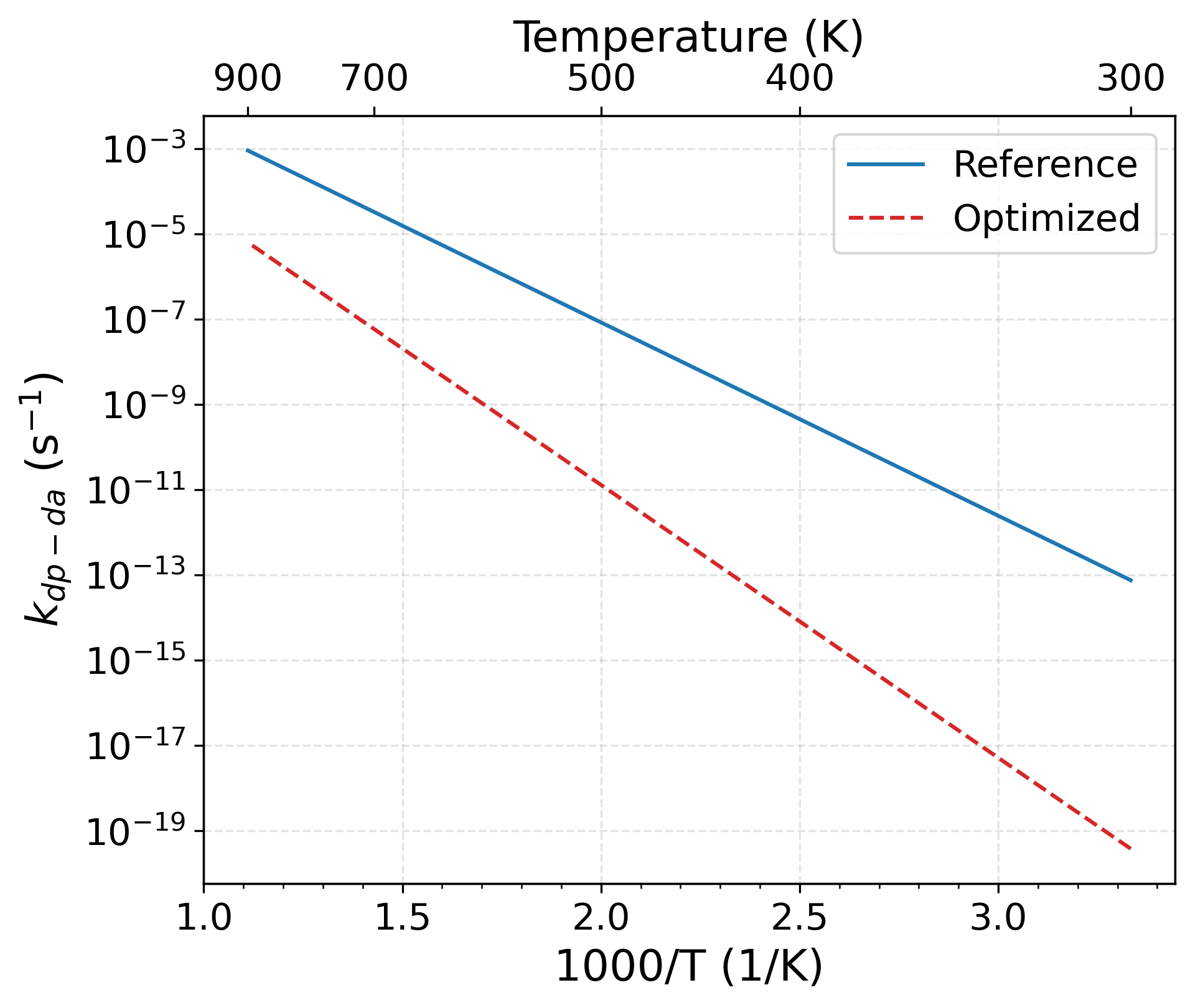}
        \caption{$k_{dp\text{-}da}(T)$}
    \end{subfigure}
    \caption{Comparison of Arrhenius temperature dependence between reference and Bayesian-optimized parameter sets for (a)~diffusivity~$D(T)$, (b)~trapping rate coefficient~$\alpha_t(T)$, (c)~detrapping rate coefficient~$\alpha_r(T)$, and (d)~annihilation rate coefficient~$k_{dp\text{-}da}(T)$.}
    \label{fig:val1:arrhenius}
\end{figure}

\Cref{fig:val1:arrhenius} compares the Arrhenius-law temperature dependence of the diffusivity~$D(T)$, trapping rate coefficient~$\alpha_t(T)$, detrapping rate coefficient~$\alpha_r(T)$, and annihilation rate coefficient~$k_{dp\text{-}da}(T)$ between the reference and optimized parameter sets over the \SIrange{300}{900}{K} TDS temperature range. The diffusivity pre-exponential factor increases by roughly one order of magnitude while the activation energy remains close to the reference (\SI{1.07}{} to \SI{1.01}{eV}). The trapping prefactor decreases by about one order of magnitude with a reduced activation energy (\SI{1.04}{} to \SI{0.82}{eV}), while the detrapping prefactor decreases by about one order of magnitude with a slightly reduced activation energy (\SI{1.19}{} to \SI{1.08}{eV}). The optimized annihilation prefactor (${\sim}\SI{83}{s^{-1}}$) remains close to the reference value (\SI{100}{s^{-1}}), with the annihilation activation energy increasing from \SI{0.9}{} to \SI{1.27}{eV}, further suppressing annihilation effects during TDS.

\begin{figure}[htbp]
    \centering
    \includegraphics[width=0.6\linewidth]{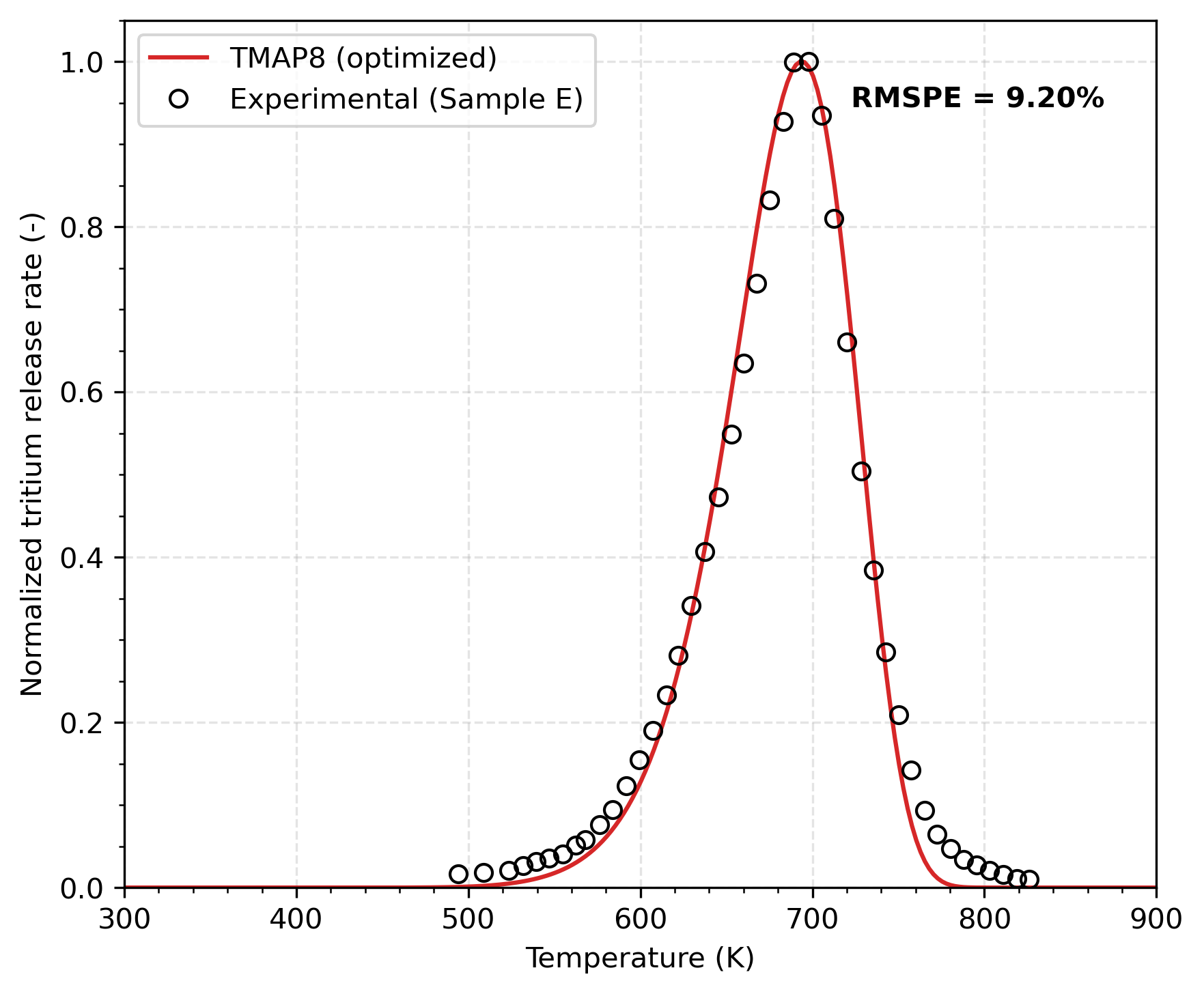}
    \caption{Comparison of TMAP8 calculation with Bayesian-optimized parameters against the experimental TDS data for Sample~E (high defect density) \cite{kobayashi2015developing}.}
    \label{fig:val1:comparison_opt}
\end{figure}

\Cref{fig:val1:comparison_opt} compares TMAP8 with the optimized parameters and the experimental data. The optimized parameters significantly reduce the RMSPE compared to the reference parameters, demonstrating improved agreement with the experimental TDS spectrum.

\subsection*{Validation case 2: Modeling the influence of thin surface oxide films on hydrogen isotope release from ion-irradiated tungsten}
\label{sec:validation:2}

\subsubsection*{Overall case description}

This validation case is based on the natural-oxide and thin-oxide experiments reported by Kremer et al. \cite{Kremer2022oxide}.
The experimental study performs thermal desorption spectroscopy (TDS) and measures deuterium release from self-irradiated tungsten samples with a natural oxide layer and with electrochemically grown oxide layers between 5 nm and 100 nm.
This case uses the same overarching model (e.g., tungsten diffusion, trapping, and surface-release formulation) to capture the deuterium release behavior of four self-irradiated tungsten samples with distinct oxygen-field configurations: natural oxide and thin oxide films of 5 nm, 10 nm, and 15 nm.
The effect of the thin oxide films on the release is discussed, with the model providing key mechanistic insights into the observed experimental behavior.
The release behavior is compared against the time-dependent experimental HD + D$_2$ and HDO + D$_2$O signals as the temperature increases.
In the present implementation, those grouped measurements are represented phenomenologically through an explicit D$_2$ release and an oxygen-gated D$_2$O release rather than through separate explicit HD and HDO transport species.

The aim of this study is to elucidate how an oxide layer affects the release of deuterium from tungsten samples and its retention. While tungsten oxidation is expected to be limited in fusion power plant conditions, it does take place in laboratory environments, which can affect laboratory observations.
Understanding oxide effects can thus help better tie laboratory experiments to performance in fusion-relevant environments, thus increasing the impact of laboratory experiments and accelerating fusion energy deployment.
Since the original paper from Kremer et al. does not propose a model to capture deuterium behavior \cite{Kremer2022oxide}, the developer and agent must develop a model from scratch for this case, unlike the workflow for the validation case presented in \cref{sec:validation:1}.

\subsubsection*{Sample history and dimensions}

The reference sample history is taken from Ref.~\cite{Kremer2022oxide}.
In the experiment, the 0.8 mm tungsten specimens are first self-irradiated, which generates a 2.3 $\mu$m thick self-damaged near-surface region.
The samples are then loaded with deuterium so that the retained inventory is concentrated in the first few micrometers of the sample.
The loading is performed at 370 K to enable deuterium mobility while minimizing defect annealing in the self-damaged region.
Once loaded, a thin oxide layer is deposited using an electrochemical process at low temperature.
The advantage of this approach compared to thermal oxidation is that the temperature remains low (e.g., room temperature), which limits deuterium transport and defect annealing.
Kremer et al. noted that electrochemically grown tungsten oxide has an amorphous structure and, therefore, differs from thermally grown oxide or natural oxide, which might affect the release behavior.

While Ref.~\cite{Kremer2022oxide} offers a wide range of data and observations for oxide-layer thicknesses reaching up to 100 nm, the current study focuses on the thinner oxide films discussed in the paper, namely a sample with a natural oxide layer and samples with oxide films that are 5 nm, 10 nm, and 15 nm thick.
The paper describes the natural oxide as being 1--2 nm thick, and the present model uses 1 nm as a representative natural-oxide value for that case.
All four desorption calculations start from the same preloaded tungsten state and follow the digitized temperature history from Fig. 6 in Ref.~\cite{Kremer2022oxide}, which shows the sample heated from approximately 296 K to approximately 1000 K over roughly 4.17 h.

\begin{figure}[h]
\centering
\includegraphics[width=0.8\textwidth]{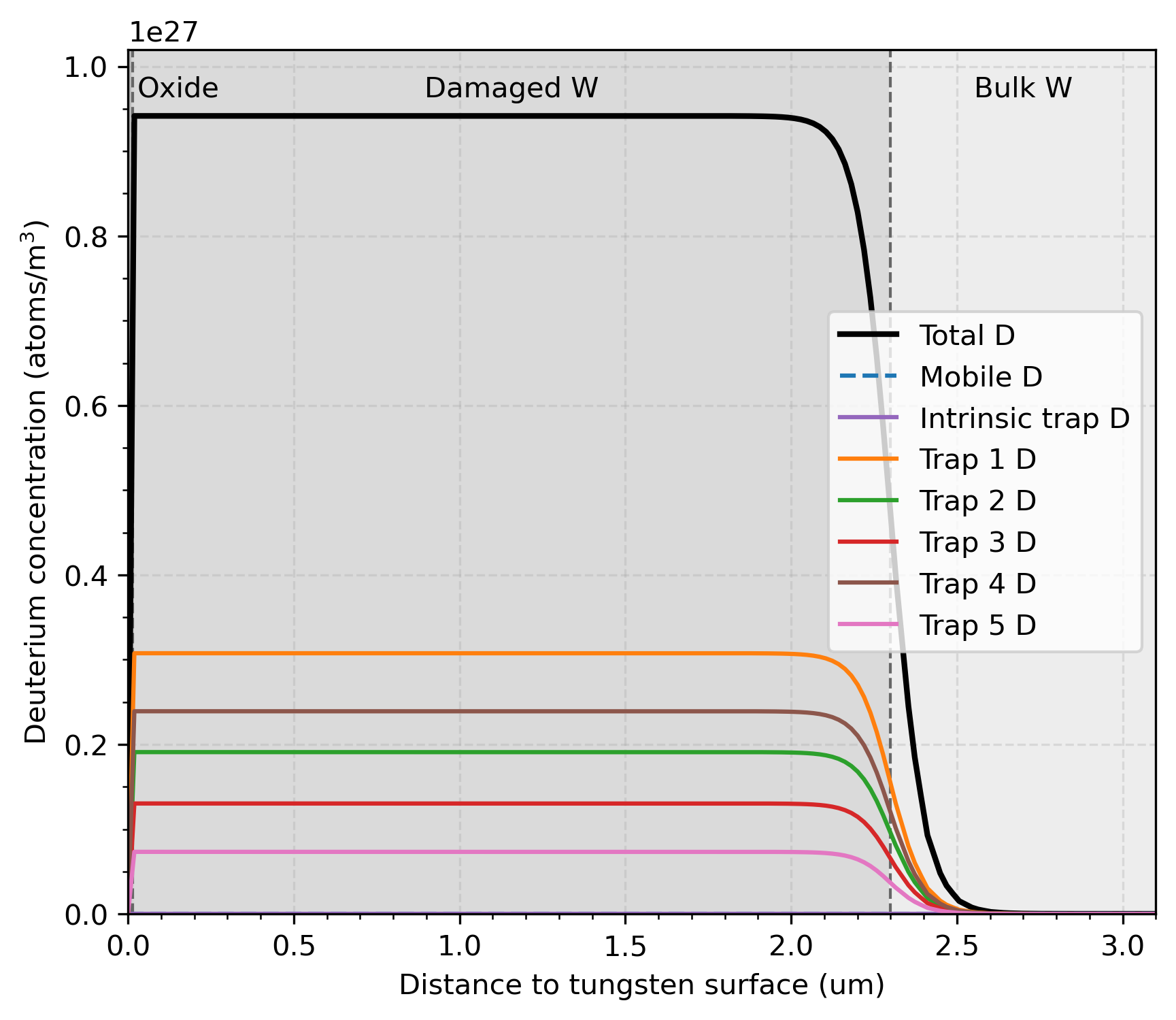}
\caption{Initial deuterium concentration profile for the desorption calculation from the 15 nm thick oxide sample. The profile shows the mobile deuterium concentration (negligible), the six trapped populations, their total, and the oxide, damaged tungsten, and bulk tungsten sections.}
\label{fig:val-2k_natural_oxide_profile}
\end{figure}

The initial deuterium profile used at the start of desorption is shown in \cref{fig:val-2k_natural_oxide_profile} for the 15 nm configuration (the same approach is used for the other samples).
The shaded regions identify the oxide, damaged tungsten, and bulk tungsten sections in the plotted depth range.
In this configuration, most of the retained inventory is placed in the irradiation-induced traps inside the damaged region, while the mobile deuterium concentration remains comparatively negligible.

\subsubsection*{Model description}

To capture the deuterium release behavior from self-irradiated tungsten with a thin oxide film, the model includes the following features:
(1) A one-dimensional geometry with an oxide layer, a self-damaged region, and the tungsten bulk, as illustrated in \cref{fig:val-2k_natural_oxide_profile}. (2) Deuterium transport involves Fickian diffusion, trapping and resolution, and surface reactions.
(3) Trapping and resolution are governed by six trap families, one intrinsic trap family and five irradiation-induced trap families. This is directly inspired by the TMAP8 validation case val-2f, which builds on the model first published by Dark et al. in Ref.~\cite{dark2024modelling}. The adapted model in val-2f validates TMAP8 based on deuterium release from self-irradiated tungsten \cite{Kadz2026}. The full set of trap site densities is adapted from val-2f so the initial areal inventory matches the prescribed preload of the current study.
(4) The density of the intrinsic trap, since it is independent of irradiation, is homogeneous in the sample. The densities of irradiation-induced traps, however, are homogeneous in the 2.3 $\mu$m thick self-damaged region, and then quickly decrease to 0 in the bulk of the sample with a transition length of 0.05 $\mu$m. Irradiation-induced trap densities also decrease to 0 in the oxide layer with a transition length of 0.25 nm (see \cref{fig:val-2k_natural_oxide_profile}).
(5) Deuterium is released either as D$_2$ or as D$_2$O by combining with an oxygen atom at the surface. The surface recombination rates of these reactions are different.
(6) The oxide layer is modeled as an additional layer on top of the self-damaged region. The transport properties of deuterium in the oxide layer remain equal to those in tungsten (e.g., same diffusivity), except that no trapping sites are present in the oxide layer. Note that the thickness of the oxygen layer does not evolve over time, even as oxygen atoms are released as D$_2$O. These simplifications are considered reasonable as the oxide layer represents only a small volume and thickness in these cases.
(7) The oxide layer is initialized with a given oxygen concentration (consistent across all cases), which is null everywhere else. The diffusivity of oxygen in the oxide layer is accounted for, but the diffusion of oxygen deeper into the tungsten sample is suppressed.
(8) The mesh is refined near the exposed surface. This is to resolve the oxide-to-damaged-tungsten and damaged-to-bulk-tungsten transition and to accurately capture the surface reactions, oxygen transport, and behavior in the self-damaged region. Deeper into the sample, the mesh is coarser to reduce computational size.

The only difference between the four configurations of interest (e.g., natural oxide and 5, 10, and 15 nm thick oxide films) is the thickness of the oxide layer and the mesh refinement area. The model formulation, other initial conditions, and all the model parameters are consistent across all cases.

The implementation is solved internally in dimensionless form, but the physical governing equations are written here for clarity. The mobile deuterium balance is
\begin{equation}
\frac{\partial C_M}{\partial t} =
\nabla \cdot \left(D_D \nabla C_M \right) -
\sum_{i \in \{intr,1,\dots,5\}} \frac{\partial C_{T_i}}{\partial t},
\end{equation}
with one trapped-species evolution equation for each trap family:
\begin{equation}
\frac{\partial C_{T_i}}{\partial t} =
\alpha_{t,i} \frac{C_{T_i}^{empty} C_M}{N} - \alpha_{r,i} C_{T_i},
\end{equation}
\begin{equation}
C_{T_i}^{empty} = C_{T_i,0} N - C_{T_i},
\end{equation}
where $C_M$ is the mobile deuterium concentration, $t$ is the time, $C_{T_i}$ is the concentration trapped in family $i$, $C_{T_i}^{empty}$ is the remaining empty trap capacity, $C_{T_i,0}$ is the fraction of host sites that can act as trap family $i$, and $N$ is the tungsten host density. $D_D$ is the deuterium diffusivity in tungsten (and tungsten oxide in this model), and $\alpha_{t,i}$ and $\alpha_{r,i}$ are the trapping and resolution rates for trapping family $i$, respectively.

The oxygen field evolves according to
\begin{equation}
\frac{\partial C_O}{\partial t} =
\nabla \cdot \left(D_O \nabla C_O \right),
\end{equation}
where $C_O$ is the oxygen concentration and $D_O$ is the oxygen diffusivity in the oxide layer.
In the current model, oxygen diffusion is masked so that it is active only inside the oxide layer.

The temperature-dependent diffusivities and trapping/detrapping rates follow Arrhenius forms:

\begin{equation}
D_D = D_{D,0} \exp \left(- \frac{E_D}{k_B T} \right)
\end{equation}
and
\begin{equation}
D_O = D_{O,0} \exp \left(- \frac{E_{D,O}}{k_B T} \right)
\end{equation}
for the diffusivities, and
\begin{equation}
\alpha_{t,i} = \alpha_{t,i,0} \exp \left(- \frac{E_D}{k_B T} \right)
\end{equation}
and
\begin{equation}
\alpha_{r,i} = \alpha_{r,i,0} \exp \left(- \frac{E_{T,i}}{k_B T} \right)
\end{equation}
for the trapping and detrapping rates.

The surface reactions represented in the model are
\begin{equation}
2D \rightarrow D_2
\end{equation}
and
\begin{equation}
2D + O \rightarrow D_2O,
\end{equation}
which give the corresponding surface fluxes
\begin{equation}
J_{D_2}^{(D\ at)} = -2 K_{r,D_2} C_M^2,
\end{equation}
\begin{equation}
J_{D_2O}^{(D\ at)} = -2 K_{r,D_2O} C_O C_M^2,
\end{equation}
and
\begin{equation}
J_O = \frac{1}{2} J_{D_2O}^{(D\ at)},
\end{equation}
where $J_{D_2}^{(D\ at)}$ and $J_{D_2O}^{(D\ at)}$ are deuterium-atom fluxes leaving the mobile-deuterium balance from $D_2$ and $D_2O$ reactions, with units of D atoms m$^{-2}$ s$^{-1}$.
Equivalently, the molecular heavy-water flux is $J_{D_2O}^{(mol)} = \frac{1}{2} J_{D_2O}^{(D\ at)}$, so the oxygen loss flux satisfies $J_O = J_{D_2O}^{(mol)}$ because one oxygen atom is consumed per released D$_2$O molecule.
While the reverse reactions, e.g., molecular dissociation at the surface, are possible, they are neglected here for simplicity due to the low partial pressure of deuterium in the gas surrounding the sample.
The surface reaction rates are defined as
\begin{equation}
K_{r,D_2} = K_{r,D_2,0} \exp \left(- \frac{E_{r,D_2}}{k_B T} \right),
\end{equation}
and
\begin{equation}
K_{r,D_2O} = K_{r,D_2O,0} \exp \left(- \frac{E_{r,D_2O}}{k_B T} \right).
\end{equation}

For numerical stability, the input files rewrite these equations in dimensionless form using the reference ratios
\begin{equation}
\hat{x} = \frac{x}{L_{\text{ref}}}, \qquad
\hat{t} = \frac{t}{t_{\text{ref}}}, \qquad
\hat{C}_M = \frac{C_M}{C_{M,\text{ref}}}, \qquad
\hat{C}_{T_i} = \frac{C_{T_i}}{C_{T_i,\text{ref}}}, \qquad
\hat{C}_O = \frac{C_O}{C_{O,\text{ref}}},
\end{equation}
with $L_{\text{ref}} = 1$ $\mu$m and $t_{\text{ref}} = 1$ s.
The corresponding dimensionless groups used in the input files are
\begin{equation}
\hat{\alpha}_{t,i} = t_{\text{ref}} \alpha_{t,i} \frac{C_{M,\text{ref}}}{N},
\end{equation}
\begin{equation}
\hat{\alpha}_{r,i} = t_{\text{ref}} \alpha_{r,i},
\end{equation}
\begin{equation}
\hat{D}_D = D_D \frac{t_{\text{ref}}}{L_{\text{ref}}^2},
\end{equation}
\begin{equation}
\hat{D}_O = D_O \frac{t_{\text{ref}}}{L_{\text{ref}}^2},
\end{equation}
\begin{equation}
\hat{K}_{r,D_2} = K_{r,D_2} \frac{C_{M,\text{ref}} t_{\text{ref}}}{L_{\text{ref}}},
\end{equation}
and
\begin{equation}
\hat{K}_{r,D_2O} = K_{r,D_2O} \frac{C_{M,\text{ref}} t_{\text{ref}}}{L_{\text{ref}}}.
\end{equation}

\subsubsection*{Case and model parameters}

The literature-based and calibrated model parameters, geometry, and sample history conditions are listed in \cref{val-2k_parameters}.
The initial oxygen concentration is derived from the reported removal of $100 \times 10^{19}$ O/m$^2$ from the first 13.5 nm of oxide in Ref.~\cite{Kremer2022oxide} and is reduced by an additional factor of 1.5 in the current calibrated model, which yields about $4.94 \times 10^{28}$ O/m$^3$.

\begin{table}[h]
\centering
\caption{Literature-based and calibrated model parameters, geometry, and sample history conditions used to investigate the effect of oxide on deuterium release.}
\label{val-2k_parameters}
\begin{tabular}{lllll}
\toprule
Parameter & Description & Value & Units & Reference \\
\midrule
$l_W$ & Tungsten thickness & 0.8 & mm & \cite{Kremer2022oxide} \\
$l_{ox}$ & Oxide thickness & 1, 5, 10, 15 & nm & \cite{Kremer2022oxide} \\
$w_{ox}$ & Oxide-to-W transition width & 0.25 & nm & Numerical resolution choice \\
$l_d$ & Self-damaged depth & 2.3 & $\mu$m & \cite{Kremer2022oxide} \\
$T_0$ & Initial desorption temperature & $\approx$ 295.775 & K & Fig. 6 in \cite{Kremer2022oxide} \\
$T_f$ & Final desorption temperature & $\approx$ 1001.408 & K & Fig. 6 in \cite{Kremer2022oxide} \\
$t_f$ & Final desorption time & 4.166 & h & Fig. 6 in \cite{Kremer2022oxide} \\
$D_{D,0}$ & Deuterium diffusivity prefactor & 1.6 $\times 10^{-7}$ & m$^2$/s & From [val-2f](val-2f.md) \\
$E_D$ & Deuterium diffusion activation energy & 0.28 & eV & \cite{Kadz2026} \\
$D_{0,O}$ & Oxygen diffusivity prefactor & 2.0 $\times 10^{-17}$ & m$^2$/s & Calibrated from Ref.~\cite{Jiang2009oxygenDiffusion} \\
$E_{D,O}$ & Oxygen diffusion activation energy & 0.45 & eV & Calibrated from Ref.~\cite{Jiang2009oxygenDiffusion} \\
$C_{O,0}$ & Initial oxygen concentration & 4.94 $\times 10^{28}$ & at/m$^3$ & Adapted from Ref.~\cite{Kremer2021oxideBarrier} \\
$w_d$ & Damaged-to-bulk W transition width & 0.05 & $\mu$m & Numerical resolution choice \\
$L_{\text{ref}}$ & Reference length  & 1 & $\mu$m & Adapted from Ref.~\cite{Kadz2026} \\
$t_{\text{ref}}$ & Reference time & 1 & s & Adapted from Ref.~\cite{Kadz2026} \\
$C_{M,\text{ref}}$ & Mobile reference concentration & 6.3222 $\times 10^{16}$ & at/m$^3$ & Adapted from Ref.~\cite{Kadz2026} \\
$C_{T,\text{ref}}^{intr}$ & Intrinsic-trap reference concentration & 6.3222 $\times 10^{17}$ & at/m$^3$ & Adapted from Ref.~\cite{Kadz2026} \\
$C_{T,\text{ref}}^{1-5}$ & Irradiation trap reference concentration & 6.3222 $\times 10^{20}$ & at/m$^3$ & Adapted from Ref.~\cite{Kadz2026} \\
$s_T$ & Uniform trap-density scale factor & 6.644848 & - & Adapted from Ref.~\cite{Kadz2026} \\
$E_{T,intr}$ & Intrinsic detrapping energy & 1.08 & eV & Adapted from Ref.~\cite{Kadz2026} \\
$E_{T,1}$ & Trap 1 detrapping energy & 1.20 & eV & Adapted from Ref.~\cite{Kadz2026} \\
$E_{T,2}$ & Trap 2 detrapping energy & 1.38 & eV & Adapted from Ref.~\cite{Kadz2026} \\
$E_{T,3}$ & Trap 3 detrapping energy & 1.65 & eV & Adapted from Ref.~\cite{Kadz2026} \\
$E_{T,4}$ & Trap 4 detrapping energy & 1.85 & eV & Adapted from Ref.~\cite{Kadz2026} \\
$E_{T,5}$ & Trap 5 detrapping energy & 2.05 & eV & Adapted from Ref.~\cite{Kadz2026} \\
$C_{T,intr,0} N$ & Intrinsic trap site density & 1.595 $\times 10^{23}$ & at/m$^3$ & Adapted from Ref.~\cite{Kadz2026} \\
$C_{T_1,0} N$ & Trap 1 site density & 3.076 $\times 10^{26}$ & at/m$^3$ & Adapted from Ref.~\cite{Kadz2026} \\
$C_{T_2,0} N$ & Trap 2 site density & 1.910 $\times 10^{26}$ & at/m$^3$ & Adapted from Ref.~\cite{Kadz2026} \\
$C_{T_3,0} N$ & Trap 3 site density & 1.304 $\times 10^{26}$ & at/m$^3$ & Adapted from Ref.~\cite{Kadz2026} \\
$C_{T_4,0} N$ & Trap 4 site density & 2.392 $\times 10^{26}$ & at/m$^3$ & Adapted from Ref.~\cite{Kadz2026} \\
$C_{T_5,0} N$ & Trap 5 site density & 7.330 $\times 10^{25}$ & at/m$^3$ & Adapted from Ref.~\cite{Kadz2026} \\
$K_{r,D_2}$ & Recombination prefactor & 3.8 $\times 10^{-16}$ & m$^4$/at/s & Adapted from Ref.~\cite{Kadz2026} \\
$E_{r,D_2}$ & Recombination activation energy & 0.34 & eV & Adapted from Ref.~\cite{Kadz2026} \\
$K_{r,D_2O}$ & D$_2$O surface-release prefactor & 3.8 $\times 10^{1}$ & m$^4$/at/s & Calibrated \\
$E_{r,D_2O}$ & D$_2$O surface-release activation energy & 2.10 & eV & Calibrated \\
\bottomrule
\end{tabular}
\end{table}

\subsubsection*{Results}

The main interpretation from Ref.~\cite{Kremer2022oxide} is that the oxide film acts both as a deuterium reservoir and as a transport barrier, delaying release as the oxide gets thicker. 
The paper also reports that chemical interaction between outgassing deuterium and the oxide begins above about 475 K, and that heavy-water release dominates above about 700 K while enough oxide remains available. 
The calibrated model developed here is evaluated against those trends as well as against the digitized TDS curves.

\begin{figure}[h]
\centering
\includegraphics[width=0.8\textwidth]{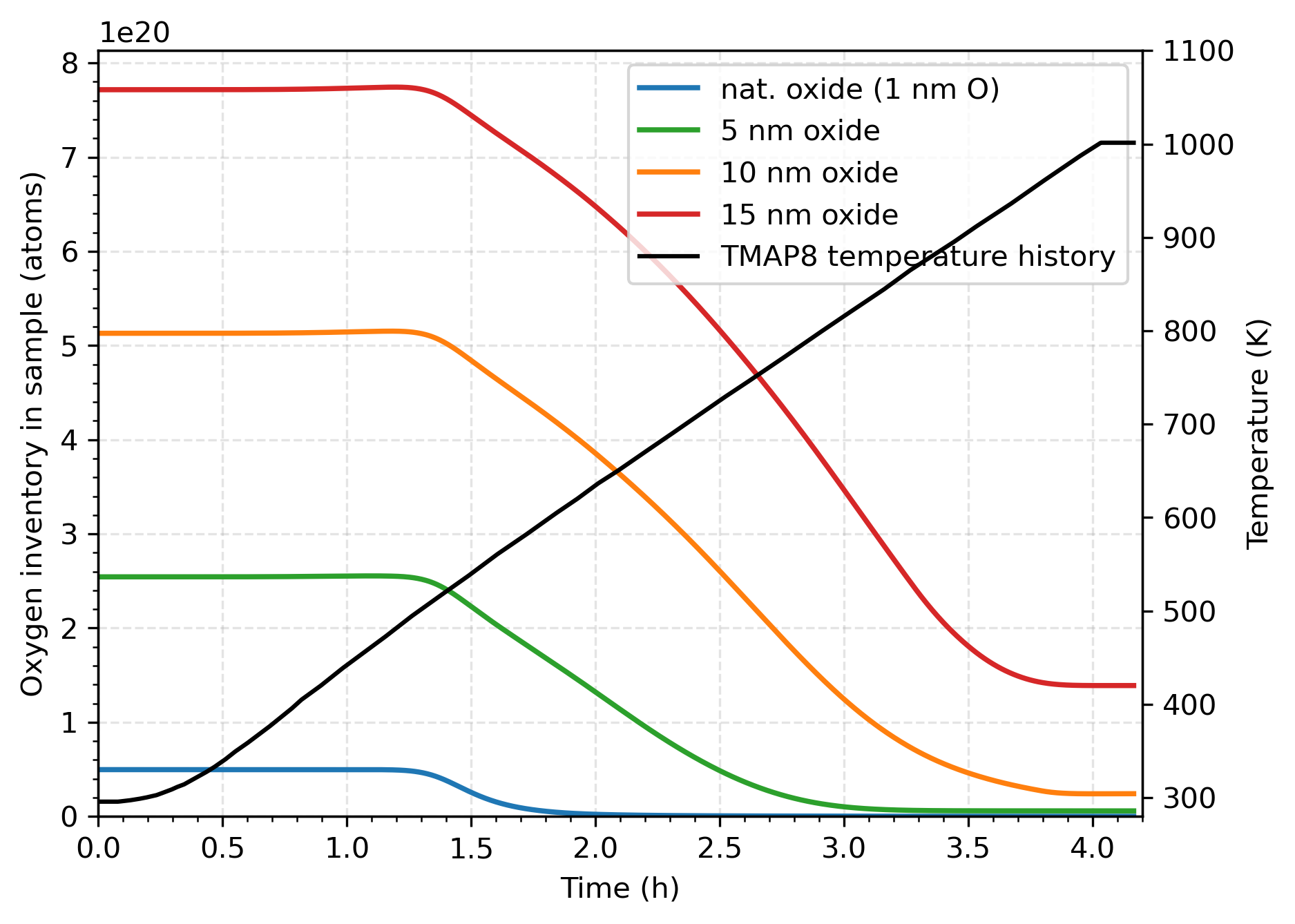}
\caption{Total oxygen inventory remaining in the sample over time.}
\label{fig:val-2k_natural_oxide_oxygen_inventory}
\end{figure}

In the case of the natural oxide shown in \cref{val-2k_natural_oxide_case_comparison}, the deuterium release is dominated by D$_2$ release with two main peaks dictated by the trapping energies.
This is consistent with the results discussed in TMAP8's val-2f case \cite{dark2024modelling,Kadz2026}.
The low release in D$_2$O form is attributed to less oxygen being available, since the 1 nm natural-oxide inventory is quickly depleted, as shown in \cref{fig:val-2k_natural_oxide_oxygen_inventory}.
The model qualitatively captures the main trends observed experimentally.
The position and magnitude of the two peaks for D$_2$ release are predicted, as well as the ratio of D$_2$ to D$_2$O release.
The main difference in trends is the short peak in D$_2$O instead of the wider peak observed experimentally.
This could be attributed to an overestimation of the oxygen availability or D$_2$O surface reaction rate at lower temperatures.

\begin{figure}[h]
\centering
\includegraphics[width=0.8\textwidth]{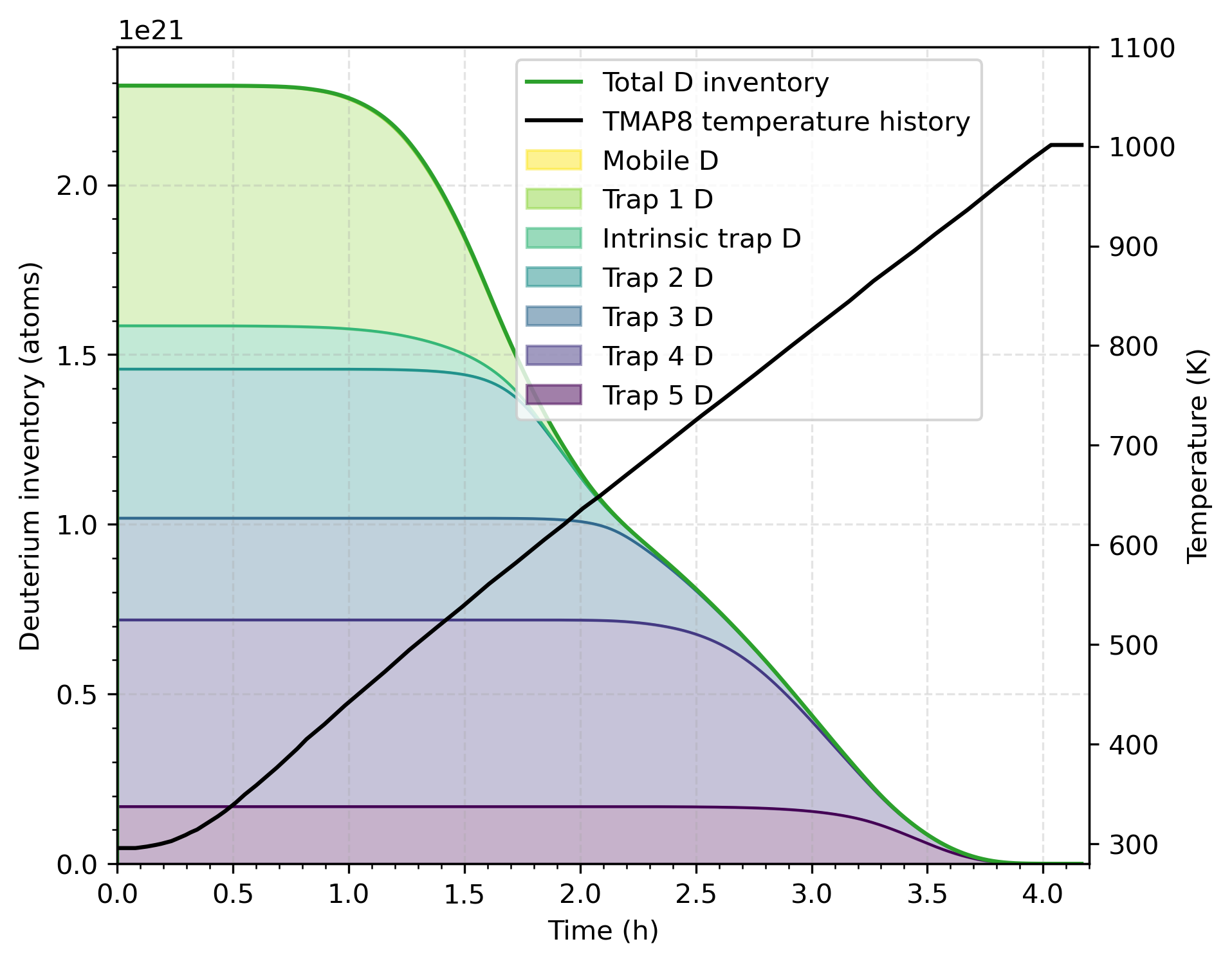}
\caption{Evolution of the mobile and trapped deuterium inventories during desorption for the 1 nm natural-oxide sample. This behavior is common to all four cases.}
\label{fig:val-2k_natural_oxide_inventory}
\end{figure}

\cref{fig:val-2k_natural_oxide_inventory} shows the evolution of the deuterium inventory as a mobile species and in each trap over time.
The deuterium release is dominated by the trapping populations, while the mobile deuterium inventory remains much smaller throughout the desorption ramp, as mobile deuterium quickly reacts at the surface of the sample.
The lower-energy traps begin to empty first as the temperature rises, followed by the deeper trap populations later in the ramp, as expected.
This behavior was common across all four cases, and the oxide thickness had no significant effect on detrapping behavior.
This was expected since the traps description was the same for all four cases.

As the oxide film thickness increases, the following trends are observable experimentally (Fig. 2 of manuscript):

\begin{itemize}
\item The ratio of D$_2$O release over D$_2$ increases as oxygen availability increases.
\item The low-temperature D$_2$ peak maintains its position, but its magnitude decreases consistently.
\item The high-temperature peak shifts to higher temperatures and its magnitude decreases, even disappearing when the oxide thickness increases from 10 nm to 15 nm.
\item The D$_2$O release increases, with a first peak aligned with the first D$_2$ peak, then a stable region, and then either a decrease in the case of the 5 nm thick oxide or another peak aligned with the second D$_2$ peak for thicker oxides. This secondary D$_2$O peak for the 10 nm oxide decreases sooner than for the 15 nm oxide.
\end{itemize}

These trends are all qualitatively captured by the calibrated model.
Furthermore, even if the model lacks a purely mechanistic description of the deuterium and oxide behavior, the simulations offer some physical insights to explain these observed trends.

Oxygen availability was found to be a key parameter during model calibration.
As the oxide thickness increases and the oxygen inventory increases (see \cref{fig:val-2k_natural_oxide_oxygen_inventory}), the ratio of D$_2$ to D$_2$O release decreases.
Then, as the oxygen inventory gets depleted, D$_2$O release naturally decreases.
This helps explain the lack of a secondary peak in D$_2$O release for the 5 nm oxide sample, as well as the thinner secondary D$_2$O peak for the 10 nm oxide sample compared with the 15 nm oxide sample.
As described in Ref.~\cite{Kremer2022oxide}, the oxide layer disappears during TDS for most cases, but some remains for the 15 nm thick oxide film case (see \cref{fig:val-2k_natural_oxide_oxygen_inventory}).
Note that while no oxide was observed after TDS for the 10 nm sample, the simulation predicts some remaining inventory, albeit only a small fraction of the initial amount. The oxide is completely gone for the natural oxide and 5 nm thickness cases in both experiments and simulations. 

\begin{figure}[h]
\centering
\includegraphics[width=0.8\textwidth]{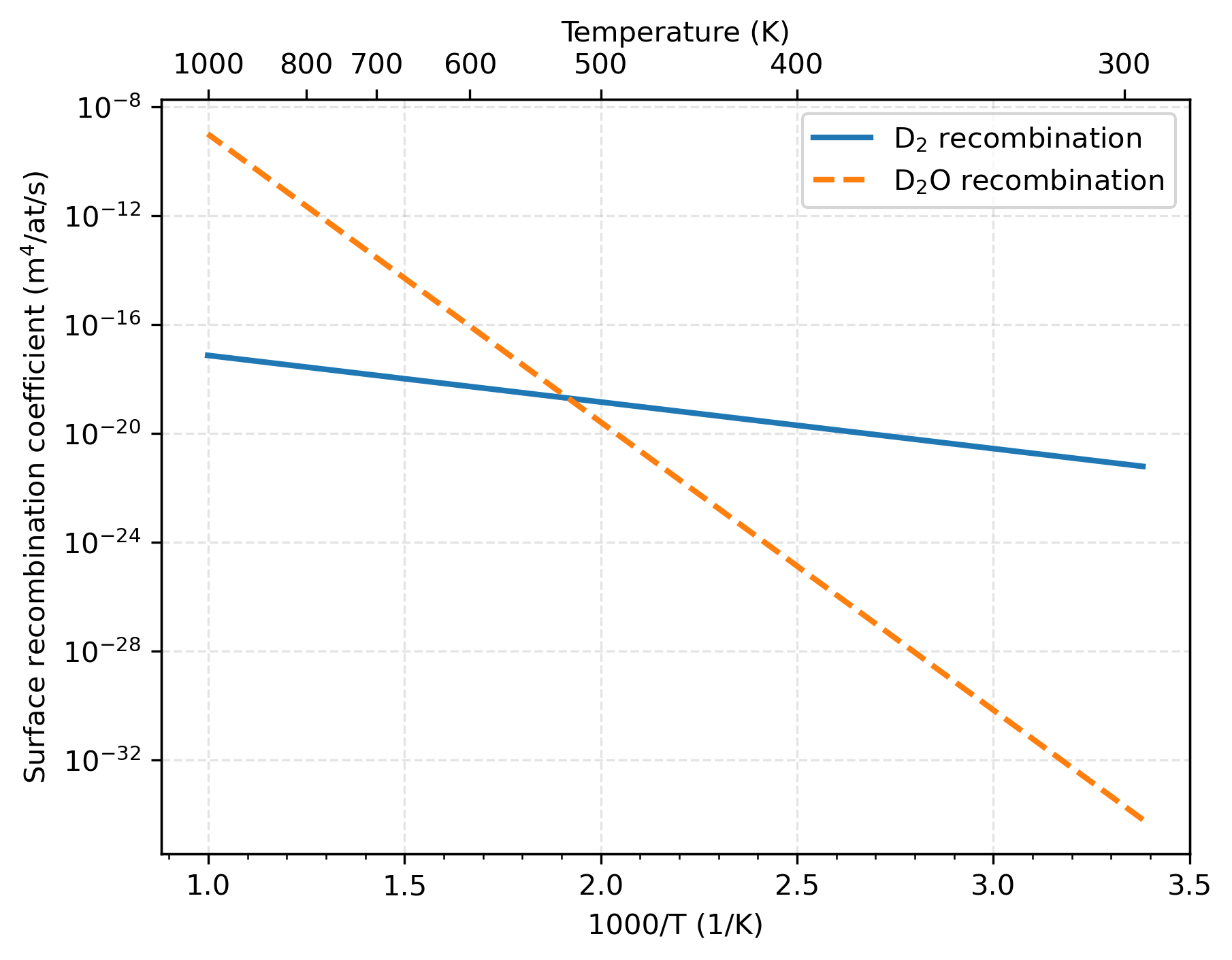}
\caption{Arrhenius-form surface recombination coefficients used for the D$_2$ and D$_2$O release reactions.}
\label{fig:val-2k_natural_oxide_recombination_rates}
\end{figure}

For oxygen to be effectively used for D$_2$O release, however, the ratio of the D$_2$ and D$_2$O surface reaction rates must be advantageous, and the oxygen diffusion in the oxide layer needs to be sufficient.
The slight delay in the onset in D$_2$O release at low temperature compared to D$_2$ release is captured by a lower D$_2$O surface reaction rate at low temperature.
However, at high temperature, the surface reaction rate of D$_2$O needs to surpass that of D$_2$ for the suppression of the secondary D$_2$ peak in favor of the secondary D$_2$O peak to be observed.
\cref{fig:val-2k_natural_oxide_recombination_rates} shows the two phenomenological surface-release coefficients over the experimental desorption temperature window.
In the calibrated parameter set, the D$_2$O release is strongly suppressed at low temperature by its larger activation energy, then rises more steeply and overtakes the D$_2$ coefficient at about 520 K, enabling the behavior discussed above.

In the case of the 15 nm thick oxide, the secondary D$_2$ peak is not completely suppressed in the current model.
However, this might be resolved with further model calibration.

\subsubsection*{Discussion and future work}

The model proposed herein uses a general formulation and consistent parameters for all four samples with different oxide thicknesses, and it qualitatively captures the main experimentally observed trends and differences between all configurations published in Ref.~\cite{Kremer2022oxide}.
By doing so, it provides key physical insights into the experimental measurements and observations.
These insights are valuable for tying laboratory observations, where tungsten oxidation often takes place, to performance in fusion power plant environments.
Using this novel model, the laboratory deuterium TDS data can be analyzed, and the effect of the oxide layer can be isolated, hence providing a model applicable to fusion energy system conditions.

This model, however, has limitations that should be addressed in future work.
The limitations discussed in Ref.~\cite{Kremer2022oxide} (e.g., electrochemically grown oxide being different from thermally grown oxide) still apply to this study.
A more thorough characterization of the oxide and a general analysis that includes different oxide structures would help generalize the current model, which currently does not differentiate between different oxide types.
In addition, the model includes other key assumptions and simplifications that could be challenged in the future to confirm the interpretation proposed in this study.
For example, the model does not capture the increased surface diffusion of deuterium, which was discussed in the original paper \citep{Kremer2022oxide} as a key release mechanism, as deuterium atoms diffuse along the sample surface to find remaining pockets of oxygen to be released as D$_2$O.
To model this, the geometry should be expanded to a two- or three-dimensional model, which is possible in TMAP8 \citep{Franklin2025,Shimada2024,Simon2022,Simon2025}.

The current study implemented the model and performed an ad hoc calibration of the model parameters based on the potential driving mechanisms of oxide evolution and deuterium detrapping, diffusion, and surface reactions.
While the experimentally observed trends are qualitatively captured by the model, the simulation results are quantitatively different from the experimental measurements.
Applying Bayesian inference to all the experimental data would allow the model to be calibrated while accounting for uncertainties related to model inadequacy, experimental error, and parameter uncertainty \citep{DHULIPALA2026102776,DHULIPALA2025155795,slaughter2023moose}.
The current oxygen diffusivity and D$_2$ and D$_2$O release parameters are, therefore, best interpreted as calibrated effective kinetics for matching the observed TDS trends rather than as a mechanistic description, which will be the goal of future work.

\end{document}